\documentclass[twocolumn]{aastex631}

\usepackage{xcolor}
\usepackage{soul}
\usepackage[most]{tcolorbox}

\newcommand{\MUSEpack}{\dataset[MUSEpack]{\doi{10.5281/zenodo.3433996}}}
\newcommand{\pyspeckit}{\texttt{pyspeckit }\citep{Ginsburg2011}}

\newcommand{\pampelmuse}{\texttt{PampelMuse }\citep{Kamann2013}}

\defcitealias{GaiaCollaboration2016b}{Gaia Collaboration et al. 2016}
\defcitealias{GaiaCollaboration2018}{Gaia Collaboration et al. 2018}
\defcitealias{GaiaCollaboration2020a}{Gaia Collaboration et al. 2020}

\shorttitle{Line-of-sight rotation of NGC~346}
\shortauthors{Zeidler et al.}
\graphicspath{{./}{figures/}}

\begin{document}

\title{The internal line-of-sight kinematics of NGC~346: the rotation of the core region}

\correspondingauthor{Peter Zeidler}
\email{zeidler@stsci.edu}

\author[0000-0002-6091-7924]{Peter Zeidler}
\affil{AURA for the European Space Agency (ESA), ESA Office, Space Telescope Science Institute, 3700 San Martin Drive, Baltimore, MD 21218, USA}

\author[0000-0003-2954-7643]{Elena Sabbi}
\affil{Space Telescope Science Institute, 3700 San Martin Drive, Baltimore, MD 21218, USA}

\author{Antonella Nota}
\affil{European Space Agency (ESA), ESA Office, Space Telescope Science Institute, 3700 San Martin Drive, Baltimore, MD 21218, USA}





\begin{abstract}

We present the stellar radial velocity analysis of the central $1 \times 1\,{\rm arcmin}$ of the young massive Small Magellanic Cloud star cluster NGC~346. Using VLT/MUSE integral field spectroscopy in combination with {\em Hubble Space Telescope} photometry we extract 103 spectra of cluster member stars suited to measure accurate line-of-sight kinematics. The cluster member stars show two distinct velocity groups at $v_1 = -3.3^{+0.3}_{-0.2}\,{\rm km}/{\rm s}$ and $v_2 = 2.6^{+0.1}_{-0.1}\,{\rm km}/{\rm s}$, relative to the systemic velocity of $(165.5 \pm 0.2)\,{\rm km}/{\rm s}$, and hint for a third group at $v_3 = 9.4^{+0.1}_{-0.1}\,{\rm km}/{\rm s}$. We show that there is neither a correlation between the velocity groups and the spatial location of the stars, nor their locus on optical color-magnitude diagrams, which makes the stellar velocity a key parameter to separate individual stellar components in such a young star cluster. Velocity group 2 shows clear rotation with $\Omega_2 = (-0.4 \pm 0.1)\,{\rm Myr}^{-1}$, corresponding to $(-4.9\pm0.7)\,{\rm km}/{\rm s}$ at radial distance of 10\,pc from the center, a possible remnant of the formation process of NGC~346 through the hierarchical collapse of the giant molecular cloud. The ionizing gas has lost any natal kinematic imprint and shows clear expansion, driven by far ultra violet fluxes and stellar winds of the numerous OB stars in the cluster center. The size of this expanding bubble and its expansion velocity of $7.9\,{\rm km}/{\rm s}$ is in excellent agreement with the estimate that the latest star formation episode occurred about two million years ago.

\end{abstract}


\section{Introduction}
\label{sec:intro}


Young stars clusters (YSCs) typically form in giant molecular clouds (GMCs) through the subsequent, hierarchical merging of smaller sub-clusters  \citep[e.g.,][]{Parker2014,Krumholz2019,Adamo2020,Dominguez2021}. This hierarchical formation process is likely to result in a total net-angular momentum of the system different from zero leading to the rotation of these YSCs, which carry the imprint of their formation process \citep[e.g.,][]{Mapelli2017,Tiongco2021}.

Recent studies indeed confirmed rotation in YSCs, particularly in R136 in the Tarantula Nebula \citep{Henault-Brunet2012} and in the $h$ and $\chi$ Persei double star cluster \citep{Dalessandro2021}. Yet, other systems like the Orion Nebula Cluster \citep[ONC,][]{Zari2019a} or Westerlund~2 \citep[Wd2,][]{Zeidler2021} show distinct kinematic groups and sub-clusters, remnants of the cluster formation process, but no rotation signature has been found. Hydrodynamic simulations of turbulent molecular clouds \citep[e.g.,][]{Mapelli2017} confirm that the cloud fragmentation process followed by gravitational collapse should almost always lead to rotation, hence should be a common feature in YSCs. Studies \citep[e.g.,][]{Kim2001,Kim2008} also show that such rotation can have significant effects on YSC by accelerating their dynamical evolution, mass segregation, and in the end their long-term survivability. Additionally, cluster evolution theories and observed rotation of Globular Clusters \citep[e.g.,][]{Fabricius2014} indicate that natal rotation of YSCs should not just be present but also be significant. Stars with a higher angular momentum, hence located in the Maxwellian tail of the velocity distribution, are more likely to escape the cluster, leading to the removal of angular momentum from the system, which in turn slows down the rotation.

To better understand the evolution of YSCs it is vital to understand why in the current sample of observations only some show rotation. This can only be accomplished by a systematic study of young systems in different environments. Such systematic observations are crucial because, despite ever more powerful computers, proper tracing of the complex dynamical evolution of YSCs from their natal GMCs to a cluster after gas expulsion remains challenging. Hence only such systematic studies of the rotation profile of YSCs with different properties (i.e., age or mass) in different environments (i.e. metallicity or gravitational potential) will eventually lead to a better understanding on star cluster evolution.

With currently existing, and even the next generation telescopes, the only places where such a systematic study is feasible are the Milky Way and the Magellanic Clouds. The longevity of the {\em Hubble Space Telescope (HST)}, the advent of large field-of-view (FoV) integral field units (IFUs) and Gaia \citepalias{GaiaCollaboration2016b,GaiaCollaboration2020a}, and more powerful computers to utilize Bayesian methods like Markov Chain Monte Carlo (MCMC) simulations and fitting are also providing us with the necessary tools to systematically access and process proper motions and radial velocities of even the most crowded star clusters \citep[e.g.,][and references therein]{Bellini2017, Kamann2018a,Groschedl2019b,Herczeg2019, Sabbi2020, Zeidler2018, Zeidler2019a, Zeidler2021,Kounkel2018, Kuhn2020}.

In this project, we study the stellar and gas kinematics of the YSC NGC~346 located in N66, the most massive star-forming region in the Small Magellanic Cloud (SMC) by focusing on the line-of-sight (LoS) velocities of the central region. At an age of $\sim 3$\,Myr \citep{Sabbi2007} NGC~346 shows a prominent pre-main-sequence \citep{Nota2006} and numerous massive O and B stars \citep[e.g.,][]{Massey1989, Walborn2000,Dufton2019}. The young stellar population is highly substructured with up to 15, mostly coeval, individual sub-clusters \citep{Sabbi2007} and a total stellar mass of $3.9 \times 10^4\,{\rm M}_\odot$ following a \citet{Salpeter1955b} mass function \citep{Sabbi2008}.

This paper is structured the following: in Sect.~\ref{sec:data} we provide an overview over the data reduction and radial velocity measurements. In Sec.~\ref{sec:RV_analysis} we take a closer look at the stellar and gas LoS velocities, while in Sect.~\ref{sec:rot} we provide a detailed analysis of the rotational profile of NGC~346. In Sec~\ref{sec:discussion} we discuss our findings followed by a brief summary in Sect.~\ref{sec:summary}.

\section{Data and data reduction}
\label{sec:data}

\begin{figure*}[htb]
\plotone{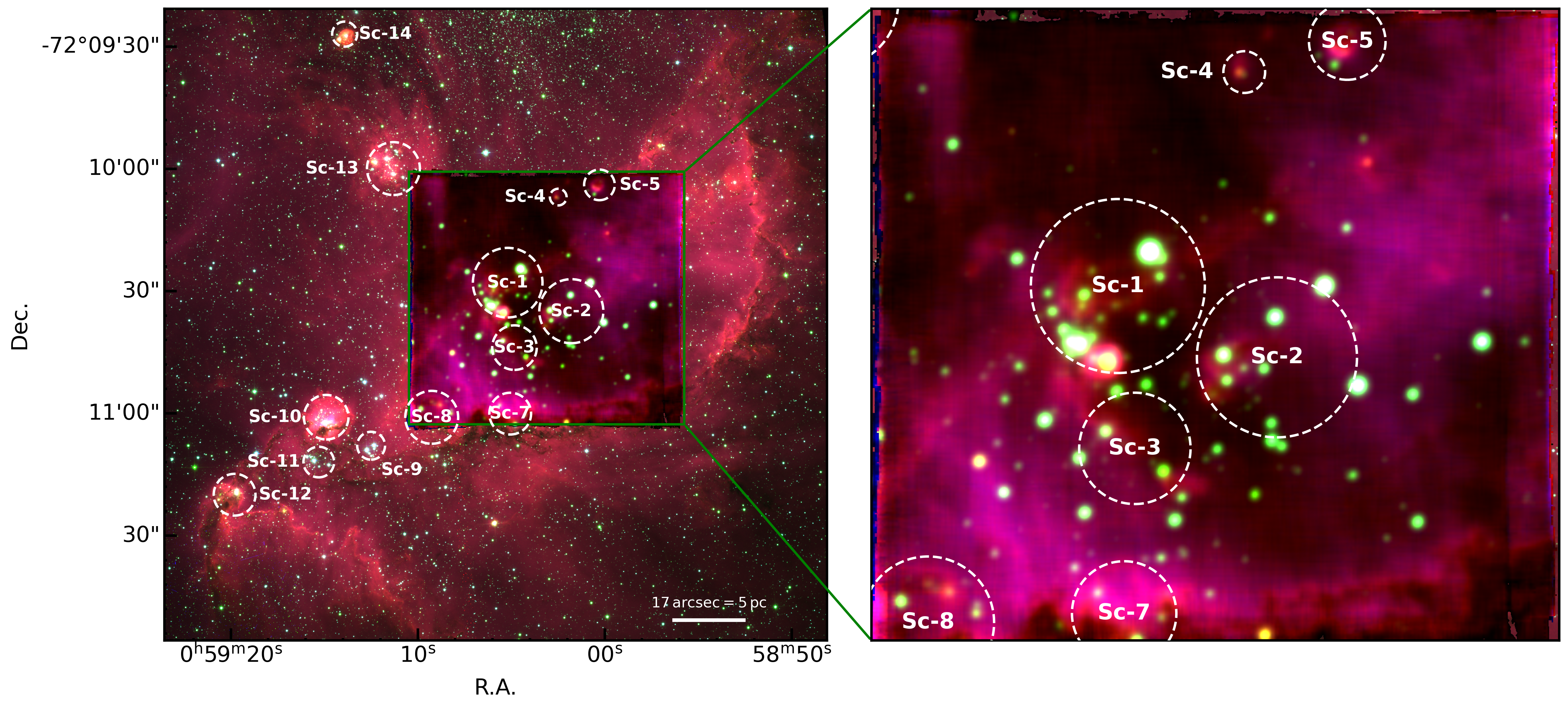}
\caption{A color-composite image of NGC~346, composed of the $F658N$ (H$\alpha$, red), $F814W$ (green), and $F555W$ (blue) {\rm HST} data. As inlay we show the MUSE dataset composed of the H$\alpha$ (red), [\ion{N}{2}]\,6582 (green), and [\ion{O}{3}]\,5007 (blue) emission lines. We also show the individual sub-clusters detected in \citet{Sabbi2007} to orient the reader. North is up, East is to the left. The right panel shows the central $1\times1{\rm arcmin}^2$ covered by MUSE.}
\label{fig:NGC346_rgb}
\end{figure*}

For this work we use data from the Multi Object Spectroscopic Explorer \citep[MUSE,][]{Bacon2010}, an integral field spectrograph mounted at UT4 of the Very Large Telescope in Chile. MUSE observes a wavelength range from 4600\,--\,9350\,\r{A} with a resolving power of $R\approx 2000$\,--\,4000 and a FoV of $1 \times 1$\,arcmin$^2$.

In the ESO observing period P98, from August 11 to 22, 2016, 48 exposures (exposure time of 315\,s each) of the central region of NGC~346 were obtained (PID: 098.D-0211(A), P.I. W.~R.~Hamann). A field rotation by 90, 180, and $270\,\deg$ between individual exposures was applied, which is the recommended strategy to mitigate detector defects. All observations were obtained in the wide-field mode without the adaptive optics system, leading to a seeing limited dataset with a DIMM seeing between $0.39"$ and $1.10"$ \footnote{When combining the individual exposures we used \texttt{weight=fwhm} in \texttt{muse\_exp\_combine}, which weights the individual exposures based on the FWHM information and is specifically implemented for data taken without AO under varying seeing conditions \citep{Weilbacher2020}}. We reduced the data using \MUSEpack~ together with version 2.8.1 of ESO's data reduction pipeline \citep{Weilbacher2012}. After a visual inspection of the individual data cubes and consulting the ESO user support, we decided to apply an additional pixel mask in regions where bright objects hampered a proper wavelength calibration. Such bright sources can cause the pipeline to fail to properly detect and fit the skylines used to calculate the wavelength offsets. Thanks to the multitude of exposures this treatment did not introduce any significant noise.

We complemented the MUSE data with optical ($F555W$, $F658N$, and $F814W$) {\em HST} photometry \citep[GO-10248, P.I.: A.~Nota,][]{Nota2006} obtained with the Wide Field Channel of the Advanced Camera for Surveys. The data reduction and the photometric catalog is described in \citet{Sabbi2007}. We transformed all coordinates to the Gaia eDR3 world coordinate system \citepalias[WCS, ][]{GaiaCollaboration2016b,GaiaCollaboration2020a} and applied a flux correction\footnote{The flux correction was applied to the individual MUSE cubes prior to their stacking.} to the MUSE data cubes relative to the $F814W$ photometry. This is needed because the absolute flux calibration of individual exposures can significantly differ depending on the observing conditions and when the standard star was observed relative to the science observations, hence this procedure will bring all exposures to an absolute flux scale prior to stacking.  As the final step we extracted the stellar spectra using \pampelmuse. For a detailed description on the data reduction process we refer to \citet{Zeidler2019a}.

\section{The radial velocity analysis}
\label{sec:RV_analysis}

\subsection{Velocity measurements}
\label{sec:vel_measurements}

\begin{figure}[htb]
\plotone{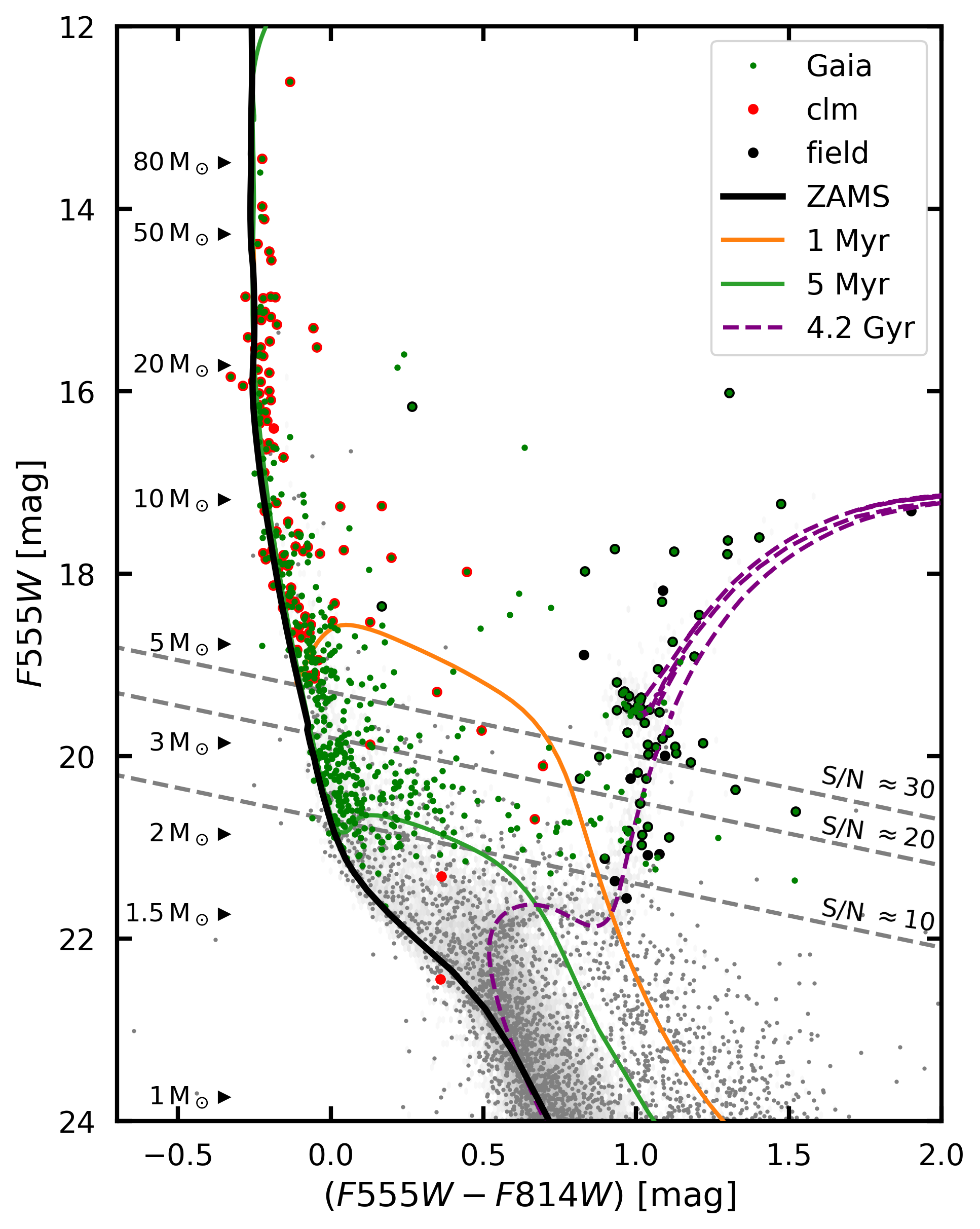}
\caption{The $F555W$ vs. $F555W-F814W$ color magnitude diagram. As reference we show all {\em HST} detected stars in grey (dots: all stars within the MUSE FoV). We marked in red (cluster members) and black (SMC field) all stars with reliable radial velocities. In green are mark all stars that are included in the Gaia eDR3 catalog. The $S/N$ limits are for the MUSE data. As orientation we also plot the zero-age main sequence, the 1 and 5\,Myr isochrone, and the 4.2\,Gyr isochrone representing the SMC field. }
\label{fig:cmd}
\end{figure}

With \pampelmuse~ we were able to extract 1005 individual spectra with $S/N \ge 5$ (Henceforth: whenever we refer to the S/N of spectra we always provide the mean S/N per spectral bin), of which 570 have a $S/N \ge 10$, and are generally suitable for radial velocity measurements \citep{Zeidler2019a,Zeidler2021}. Compared to Globular Clusters, YSCs have a highly variable background due to the large amounts of remaining ionized gas within the cluster. This gas emission does not only vary with location but also with wavelength, which complicates the local, wavelength dependent background subtraction, hence the extraction of clean spectra. We used \MUSEpack~ and the following steps to derive reliable stellar radial velocity measurements without using a spectral template catalog:

\begin{enumerate}
    \item A visual inspection of all the extracted spectra is performed to verify that the local background is correctly subtracted by \pampelmuse~, discard obviously contaminated spectra, and thus obtain a clean sample.
    \item Individual templates around each absorption line are created to measure the radial velocity shift of each extracted spectra using only the core of each line. The fitting step using pPXF \citep{Cappellari2004,Cappellari2017} is typically repeated 10,000 times and for each iteration the uncertainties of the spectrum are reordered randomly. The per line derived radial velocities are compared with each other, to systematically and objectively remove ``odd'' line profiles that might still be affected by gas emission, and to discard the spectra if the radial velocities deviate too much from each other.
    \item Last but not least, all remaining lines per star are now used together to determine its radial velocity using the same method as in the previous step with a repetition of 20,000 times. The resulting Gaussian distribution gives the radial velocity of the star (mean) and its uncertainty ($1\sigma$).
\end{enumerate}

Depending on the stellar spectral type we used the following stellar absorption lines: \ion{He}{1}\,4922, 5876, 6678, 7065, \ion{He}{2}\,4685, 5412, \ion{Mg}{1}\,5167, 5172, 5183, and \ion{Ca}{2}\,8498, 8542, 8662. The reliability of this method is discussed in \citet{Zeidler2019a}. 

\begin{figure}[htb]
\plotone{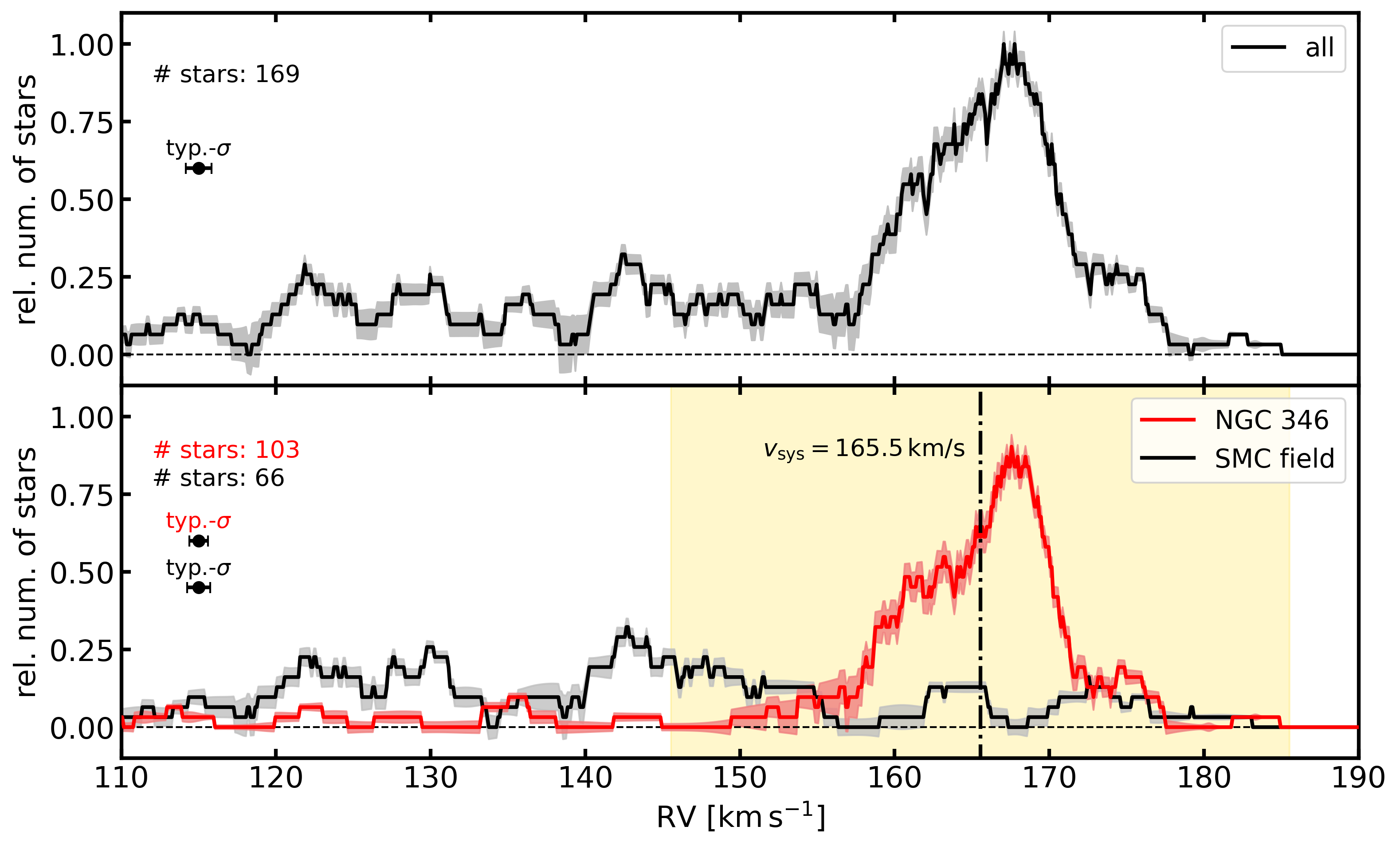}
\caption{The radial velocity distribution of all 169 stars (top panel) and the cluster member stars (red) and SMC field stars (black) in the bottom panel. We used the binning technique described in \citet{Zeidler2021}. The typical uncertainty (per star) for the cluster members and field stars are $1.6\,{\rm km}/{\rm s}$ and $2.0\,{\rm km}/{\rm s}$, respectively. The yellow region marks the $v_{\rm sys}$ by $\pm 20\,{\rm km}/{\rm s}$ used for the analysis.}
\label{fig:rv_dist}
\end{figure}

Given the severe crowding and the large amounts of remaining gas and dust, we derived reliable radial velocities for 169 stars. We applied a color selection in the $F555W$ vs. $F555W-F814W$ color-magnitude diagram (CMD, see Fig.~\ref{fig:cmd}) combined with a categorization of early-type stars (showing \ion{H}{1}, \ion{He}{2}, or broad hydrogen features, indicating an O, B, or A-type star) and late type stars (showing metal features, i.e, \ion{Ca}{2}-Triplet and \ion{Mg}{1}-Triplet to distinguish likely NGC~346 member stars from  those belonging to the SMC field. In total this selection yields 103 cluster member stars (see Tab.~\ref{tab:clm_stars} for a complete list) and 66 SMC field stars. The radial velocity profile of both groups are shown in Fig.~\ref{fig:rv_dist}. The median velocity of the NGC~346 members is $(165.5 \pm 0.2)\,{\rm km}/{\rm s}$, which we use as the systemic velocity of the cluster member stars, $v_{\rm sys}$, henceforth unless specified otherwise. This velocity is in excellent agreement with the findings of \citet{Niemela1986} who derived a mean velocity of $(163 \pm 4)\,{\rm km}/{\rm s}$ for 58 massive NGC~346 cluster member stars, and \citet{Evans2008} who measured a mean velocity of $(167.4 \pm 0.2)\,{\rm km}/{\rm s}$ and a dispersion of $33.43\,{\rm km}/{\rm s}$ for a sample of mainly early-type stars throughout the SMC bar.

For the further analysis we will only consider stars that do not exceed $v_{\rm sys}$ by $\pm 20\,{\rm km}/{\rm s}$, which are 87 of the 103 stars (see Fig.~\ref{fig:rv_dist}). The remaining 16 stars are considered as possible runaway candidates, which we will analyze in a future work\footnote{$\pm 20\,{\rm km}/{\rm s}$ was chosen because this includes the main velocity peak.}

The CMD in Fig.~\ref{fig:cmd} shows four cluster member stars (corresponding to stars 2, 61, 90, and 98 in Tab.~\ref{tab:clm_stars}) that are significantly fainter ($F555W > 20$\,mag) than the rest of the MUSE detected sources. All but star 90 are located in a relatively sparsely populated region outside the central cluster with a low background contamination, allowing us to extract clean spectra from the data cubes despite the lower signal. Stars 2 and 98 have significantly different velocities than the bulk of the cluster stars, hence these stars are probably not members of NGC~346. Given that any further analysis only includes stars with $\pm20\,{\rm km}/{\rm s}$ around the systemic velocity these four stars are excluded.

\begin{figure}[htb]
\plotone{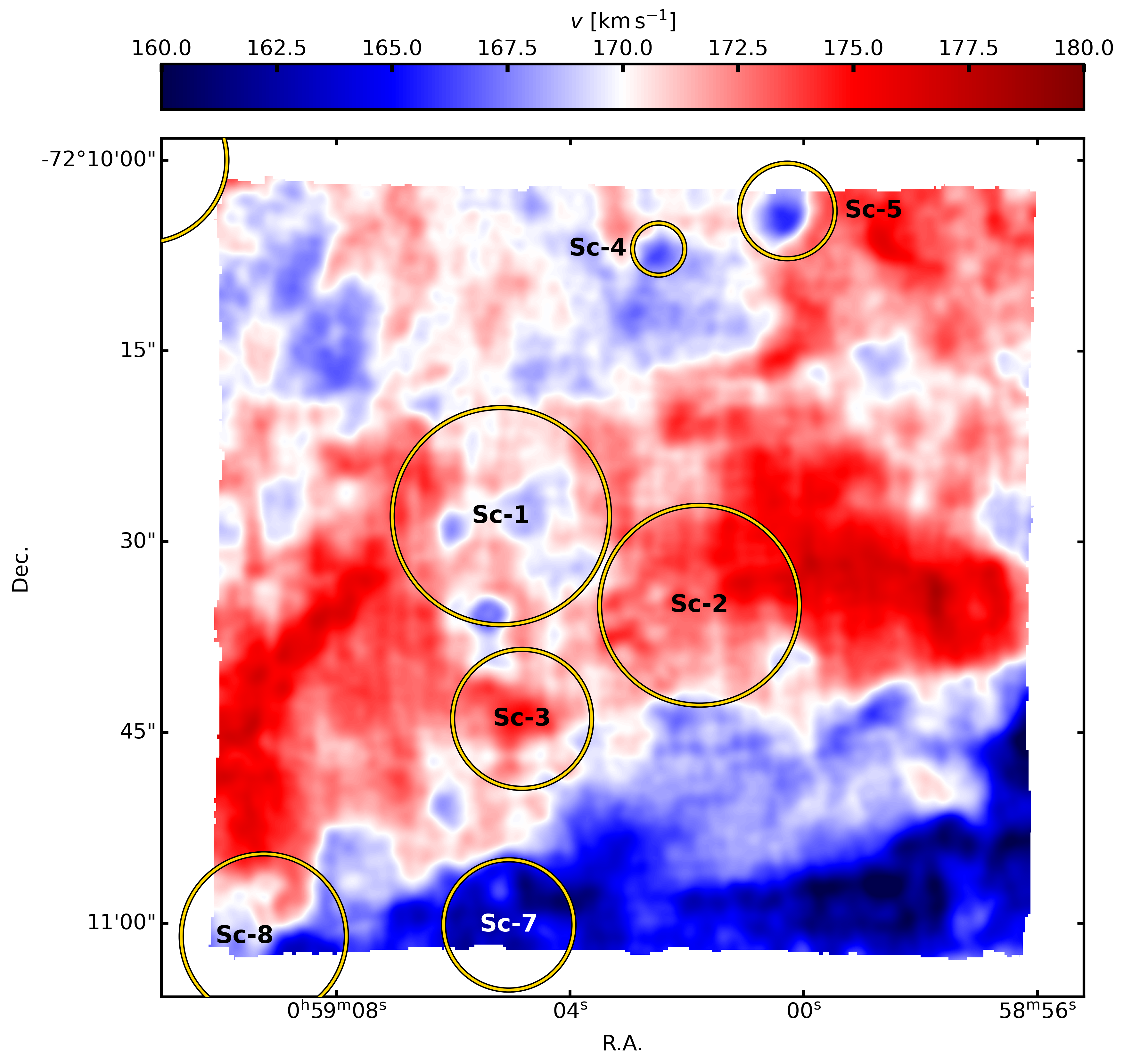}
\caption{The gas velocity map. To guide the reader we indicated the \citet{Sabbi2007} sub-clusters.}
\label{fig:rv_gas}
\end{figure}

To determine the kinematics of the ionized gas we applied the same method as, e.g., \citet{McLeod2015} and \citet{Zeidler2021} using strong gas emission lines, specifically H$\alpha$, \ion{N}{2}\,6549, 6585, and \ion{S}{2}\,6718, 6732, processed with \pyspeckit\footnote{For each spectral pixel, all emission lines are combined to one line in velocity space to measure the radial velocity. This is possible because gas emission lines are typically narrower than the MUSE dispersion, hence have a comparable shape driven only by the very stable line spread function.}. To automatically remove unreliable velocity measurements (i.e., because of stellar contamination) we masked all pixels that exceed the normalized velocity map by $5\,\sigma$ within a $32\times32$ pixel window. The window size is driven by the PSF and a visual inspection of the mask itself. Subsequently, all masked pixels were linearly interpolated. Last but not least, the velocity map was convolved with a 2D Gaussian with a width of 0.8'' representing the mean seeing. The gas velocity map is shown in Fig.~\ref{fig:rv_gas}.

\subsection{The stellar velocity profile}
\label{sec:stellar_vel_profile}

The velocity profile of the cluster member stars (see Fig.~\ref{fig:rv_dist}) has a non-Gaussian shape similar to Wd2 \citep{Zeidler2021}, indicating multiple kinematic groups. Based on the shape of the velocity profile, we use Bayesian inference to fit one or a combination of two or three Gaussians with a common offset, and the Akaike information criterion \citep[AIC,][]{Akaike1974} and the Bayesian information criterion \citep[BIC,][]{Schwarz1978} to determine the best fitting model avoiding over-fitting\footnote{The AIC (BIC) for one two and three Gaussians is 2257 (2272), 1997 (2024), and 1454 (1492), respectively, favoring the latter solution.}. The best fitting model is a combination of three Gaussians. The detailed results are shown in Tab.~\ref{tab:rv_stars}. Additionally, we simulate 5000 realizations of a single peak velocity distribution to determine the probability of the three groups being the result of small number statistics. Although in only 5.2\% of all cases a three peak solution converged, hence we can conclude at 95\% confidence that the three groups are real. Group 3 (red) only contains 5 stars, thus we will conservatively discard it from the following discussion. A complex velocity field was recently found also in a recent proper-motion study \citep{Sabbi2022}. We will discuss their results in context with ours later in this paper.

\begin{deluxetable}{crrrc}[htb]
	\tablecaption{The stellar kinematic model \label{tab:rv_stars}}
	\tablehead{\multicolumn{1}{c}{vel. group} & \multicolumn{1}{c}{$v$} & \multicolumn{1}{c}{$\sigma$} & \multicolumn{1}{c}{$n_\star$} & \multicolumn{1}{c}{color} \\
	\multicolumn{1}{c}{} & \multicolumn{2}{c}{$({\rm km}/{\rm s}$)} & \multicolumn{1}{c}{} & \multicolumn{1}{c}{}
		}
	\startdata
	 $v_1$  &  $-3.3^{+0.3}_{-0.2}$ & $3.0^{+0.2}_{-0.2}$  &  28 & blue \\
	 $v_2$  &  $2.6^{+0.1}_{-0.1}$ & $2.0^{+0.1}_{-0.1}$  &  33 & green\\
	 $v_3$  &  $9.4^{+0.1}_{-0.1}$ & $1.3^{+0.1}_{-0.1}$  &  5 & red \\
	\enddata
	\tablecomments{The best-fitting model for the stellar radial velocity distribution. In column 1 we present the group designation, in column 2 and 3 the mean velocities and their dispersions, column 4 shows the number of uniquely identified stars within $1\sigma$ of the dispersion, and column 5 shows the color that is used for all further plots. All velocities are relative to $v_{\rm sys} = (165.5\pm0.2)\,{\rm km}/{\rm s}$.}
\end{deluxetable}

\begin{figure}[htb]
\plotone{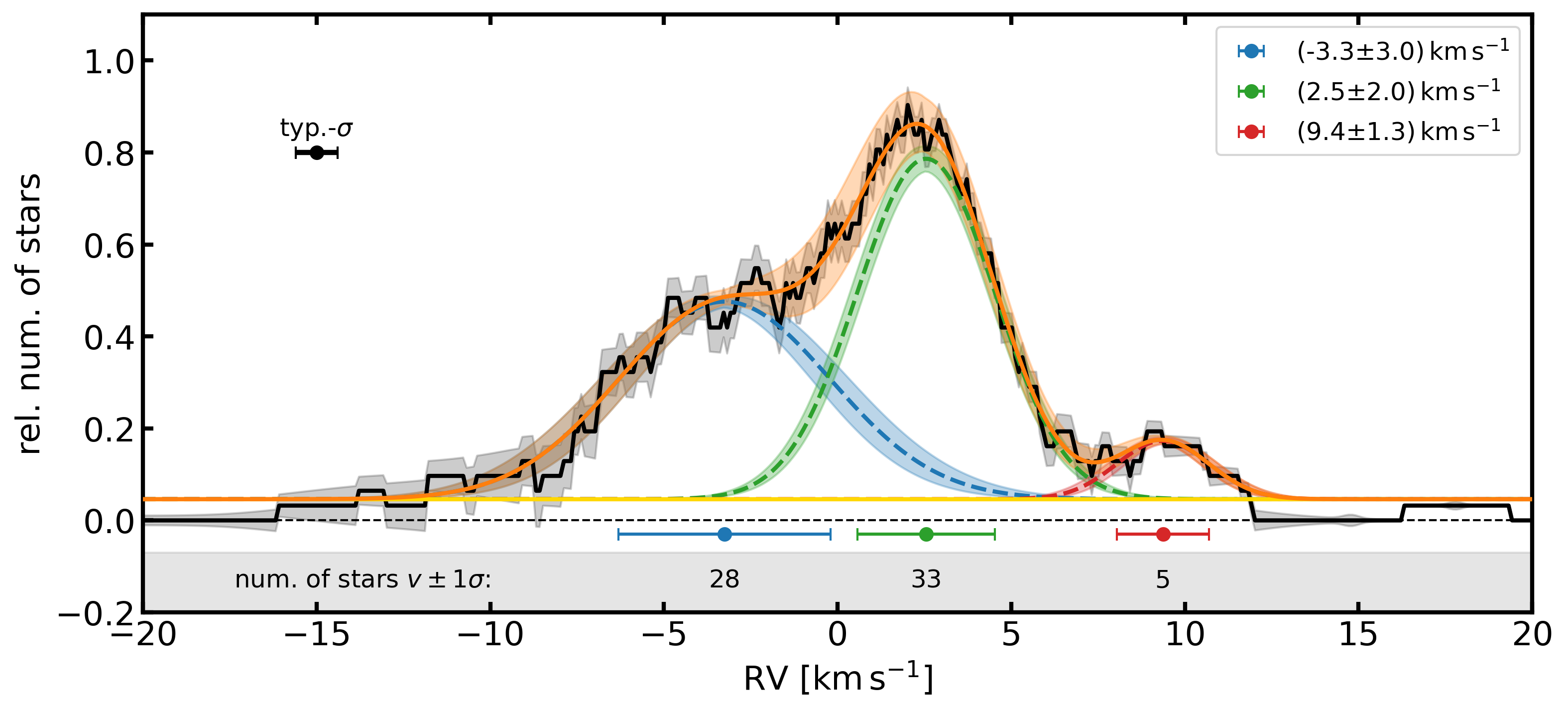}
\caption{The best-fitting model (orange) of the radial velocity distribution of the cluster-member stars. The three individual Gaussians are plotted in blue ($(-3.3 \pm 3.1)\,{\rm km}/{\rm s}$), green ($(2.6 \pm 2.0)\,{\rm km}/{\rm s}$), and red ($(9.4 \pm 1.3)\,{\rm km}/{\rm s}$). All velocities are relative to $v_{\rm sys} = (165.5\pm0.2)\,{\rm km}/{\rm s}$.}
\label{fig:rv_mod}
\end{figure}

\subsection{The gas velocity profile}
\label{sec:gas_vel_profile}

\begin{figure}[htb]
\plotone{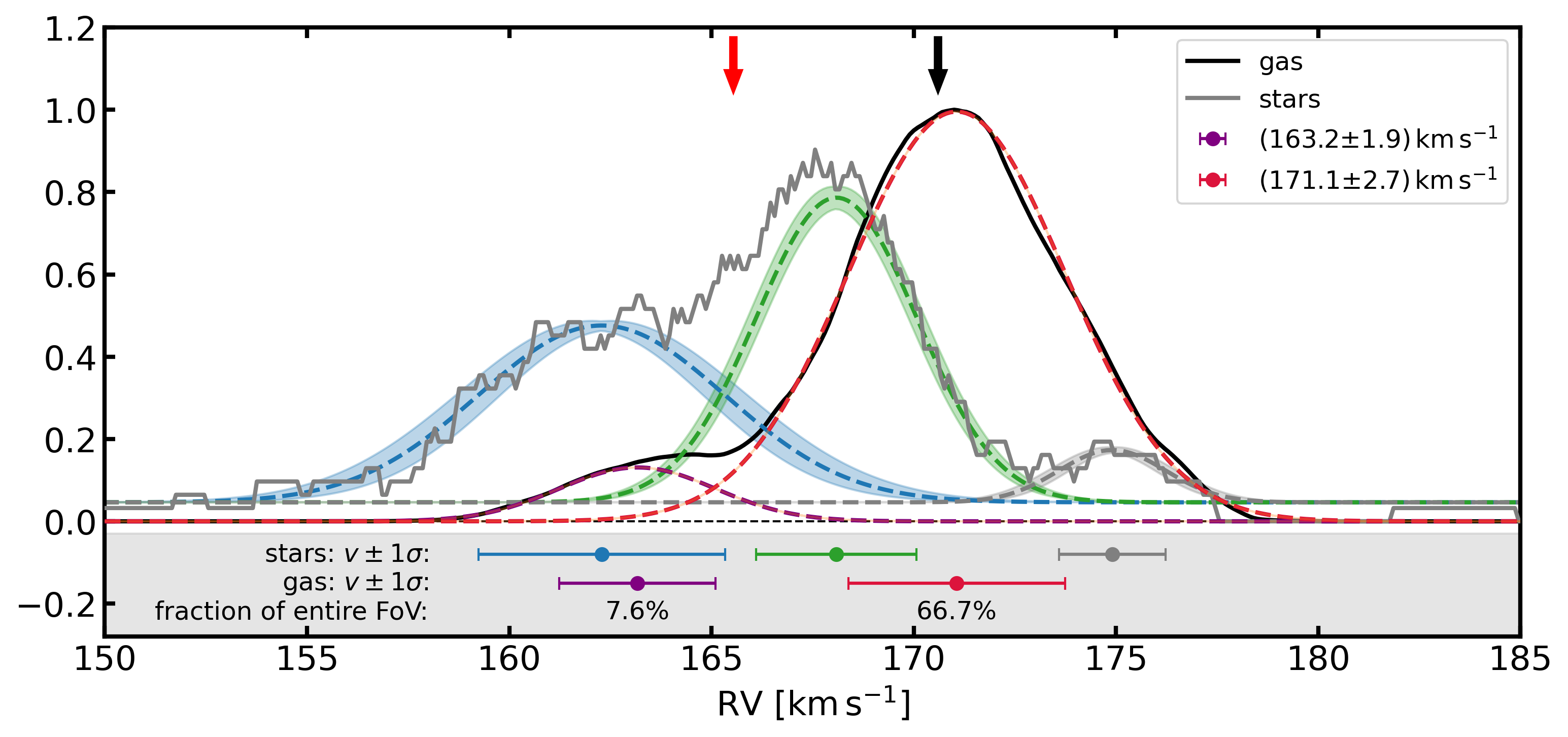}
\caption{The gas velocity profile (black curve) of the entire MUSE FoV. The two gas velocity groups are shown in purple, $v^g_1 = (163.2 \pm 1.9)\,{\rm km}/{\rm s}$, and red, $v^g_2 = (171.1 \pm 2.7)\,{\rm km}/{\rm s}$, with a total area fraction of 7.6\% and 66.7\%, respectively. We also show the distribution of the cluster stars for reference (similar to Fig.~\ref{fig:rv_mod}). We indicate the median gas velocity and the stellar systemic velocity with a black and red arrow, respectively.}
\label{fig:rv_mod_gas}
\end{figure}

To analyze the kinematics of the gas we created a velocity histogram similar to Fig.~\ref{fig:rv_mod} from the 2D velocity map (see Fig.~\ref{fig:rv_gas}). Compared to 103 stars we have over 1.5 million data points. To obtain a comparable histogram (Fig.~\ref{fig:rv_mod_gas}) we are using the same method as for the stars. This distribution clearly shows two peaks, which we fitted in the same manner as the stellar distribution using MCMC and a two Gaussian composite model. The mean velocities of the two peaks (named gas velocity groups 1 and 2 henceforth) are $v^g_1 = 163.2\,{\rm km}/{\rm s}$ and $v^g_2 = 171.1\,{\rm km}/{\rm s}$ with a dispersion of $\sigma^g_1 = 1.9\,{\rm km}/{\rm s}$ and $\sigma^g_2 = 2.7\,{\rm km}/{\rm s}$, respectively. Due to the high number of data points the statistical uncertainty is negligible. 7.6\% and 66.7\% of the entire region are located within $1\sigma$ of the dispersion of the gas velocity groups 1 and 2, respectively.

\begin{figure*}[htb]
\plotone{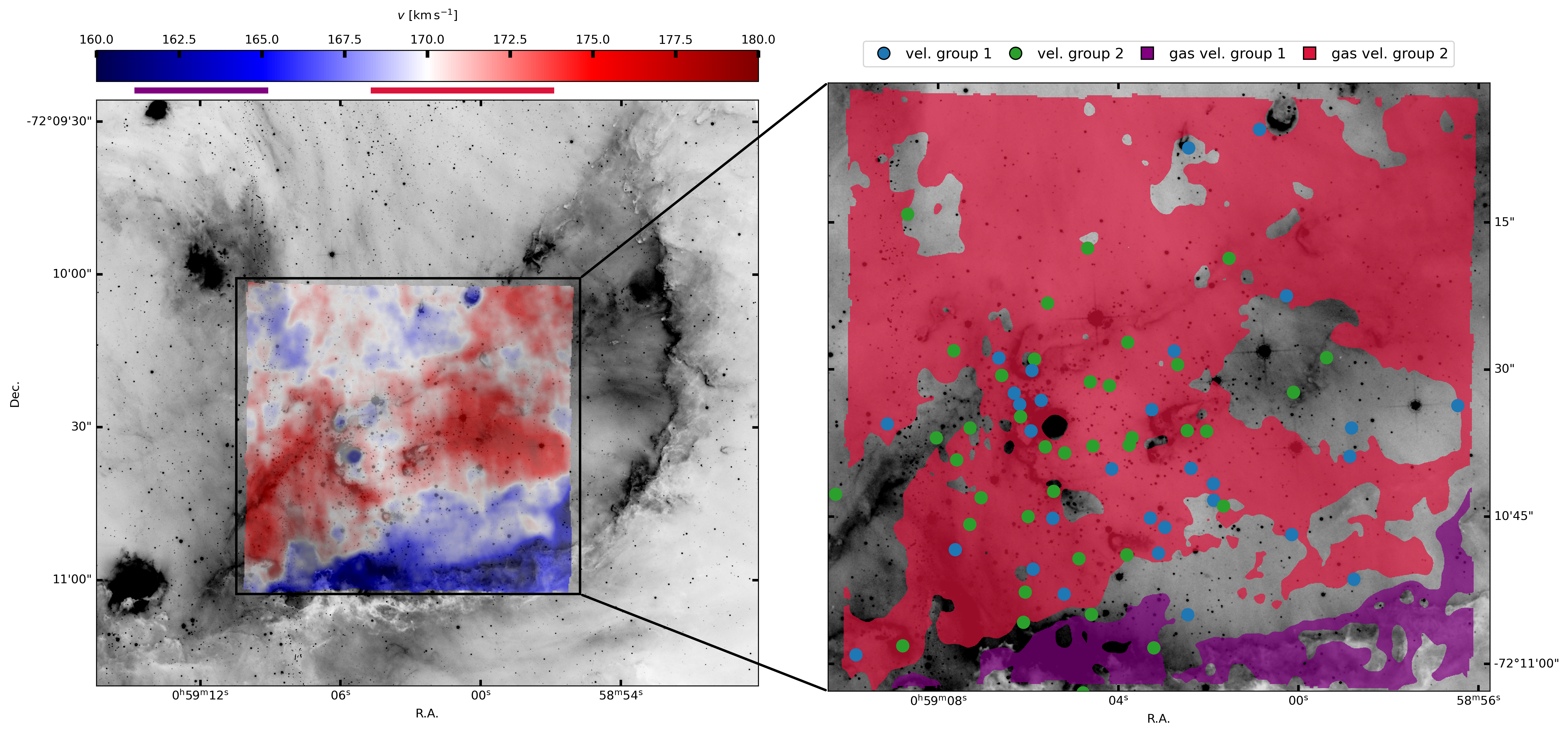}
\caption{The left panel shows the gas velocity map plotted over the {\em HST} H$\alpha$ image of NGC~346. We marked the range of both gas velocity groups underneath the color bar. The right panel shows the spatial location of both gas velocity groups as well as the stars of velocity groups 1 and 2.}
\label{fig:gas_groups_map}
\end{figure*}

In Fig.~\ref{fig:rv_mod_gas} one can see that stars of group 2, $v_2 = 168.1^{+0.1}_{-0.1}\,{\rm km}/{\rm s}$ are located between the two gas velocity groups, while the gas velocity group 1 and the stars of group 1, $v_1 = 162.3^{+0.3}_{-0.2}\,{\rm km}/{\rm s}$, are fully overlapping. When looking at the spatial distribution of both gas velocity groups (Fig.~\ref{fig:gas_groups_map}) one can clearly see that group 2 covers most of the cluster region, while group 1 covers the gas ridge to the South. There is no spatial correlation between the gas and the stellar velocity groups.

\section{The rotational profile of NGC~346}
\label{sec:rot}

\begin{figure}[htb]
\plotone{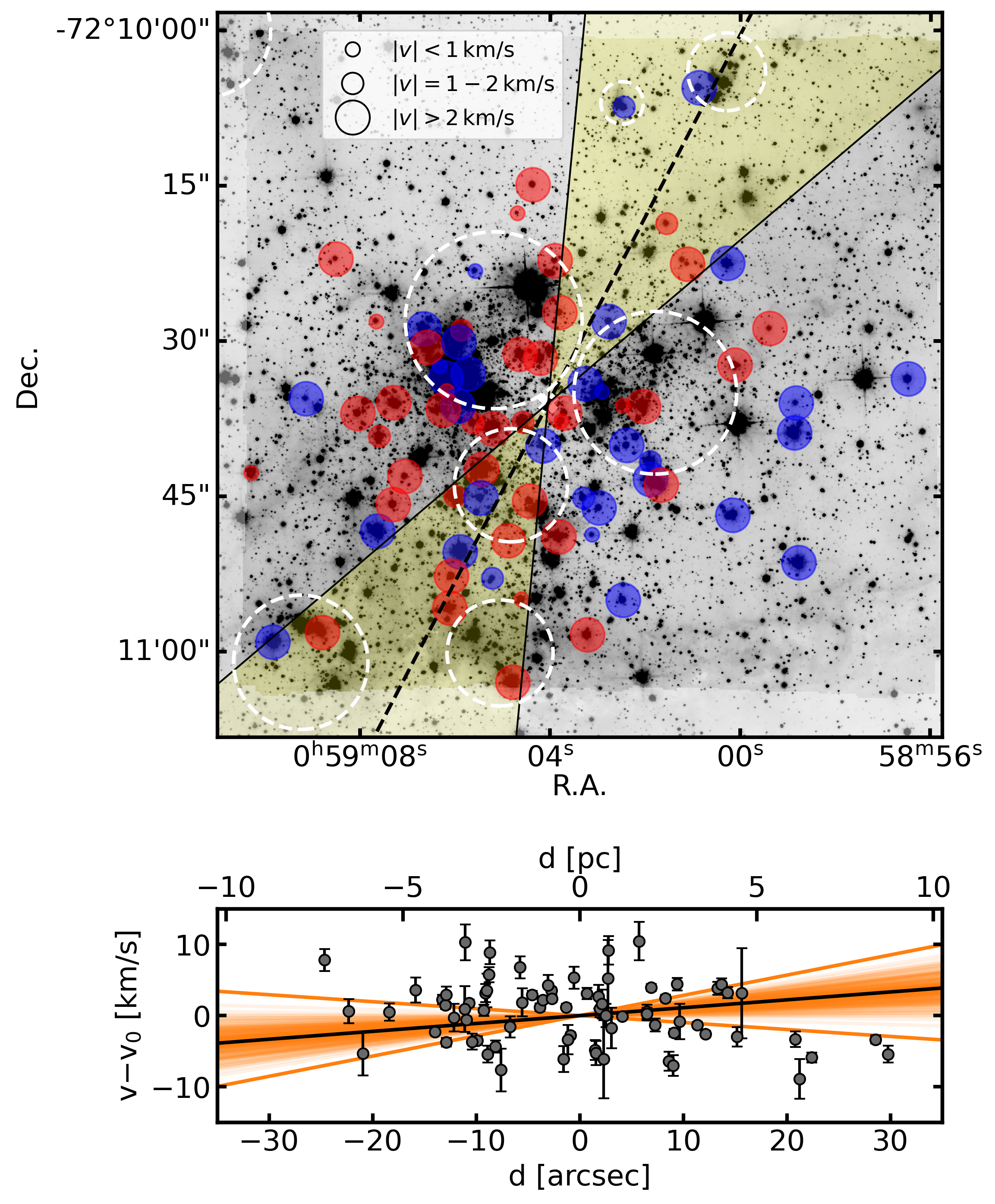}
\caption{The best-fitting rotation model to all cluster member stars. The top panel shows whether a star is red or blue shifted relative to the mean velocity. The marker size indicates the magnitude of the velocity in three bins: $|v|<1\,{\rm km}/{\rm s}$, $1\,{\rm km}/{\rm s} \le |v| < 2\,{\rm km}/{\rm s}$, $|v| \ge 2\,{\rm km}/{\rm s}$). The dashed line is the best fit rotation axis with the error cone shown in yellow. The sub-clusters from \citet{Sabbi2007} are indicated as white dashed circles to guide the reader. The bottom panel shows the stellar radial velocity relative to the distance to the rotation axis. The black line is the best-fit model while in orange we show a randomized sub-sample of all the MCMC solutions. The thicker line represents the most extreme solution of the fit.}
\label{fig:rot_all}
\end{figure}

A visual inspection of the location of all cluster member stars, color-coded by their radial velocity (see Fig.~\ref{fig:rot_all}) hints for a slight over abundance of blue-shifted stars toward the Southeast, while the number of red-shifted stars appears to be higher toward the Northwest. Due to the absence of a spatial correlation between groups 1 and 2 (see previous section) we further investigate whether this could be related to rotation, similar to what is seen in R136 \citep{Henault-Brunet2012}, or if this is a result of small number statistics. We chose one of the simplest models, a solid body rotator, defined by the angular velocity $\Omega$. The observed velocity, $v_{\rm rot}$, at any given point in space only depends on the distance, $d$, to the rotation axis. For this model and an unknown inclination $i$ of the system, the observed LoS velocity of a star $j$ is: $[v_{\rm rot}\,\sin{i}]_j = \Omega\,d_j$. The reason behind the choice of such a rather simple model over more sophisticated, physically motivated one \citep[e.g.,][]{Lynden-Bell1967,Henault-Brunet2012,Kamann2020a} the relatively small, incomplete sample of stars and the observations do not cover the full extend of NGC~346, hence the spatial profile is most likely truncated.

We considered two scenarios: 1) All stars are following the same rotation profile and 2) the two detected velocity groups are the result of two kinematically different stellar populations. For scenario 1 we fit one rotation profile to all stars. For scenario 2 we combined two rotation profiles, while we gave each star a probability to be part of either group. This probability is given by the model of the velocity profile of Sect.~\ref{sec:RV_analysis}, taking into account the individual measurement uncertainty\footnote{To not skew the plots toward the blue or red-shifted wing of each of distribution we apply an additional criterion that the probability of each star had to exceed $1\,\sigma$}. For both scenarios we introduce the position angle $\Phi$ of the rotation axis, which is defined counter-clockwise with $\Phi=0^\circ$ pointing North, hence, $\Phi=90^\circ$ pointing East and for $0^\circ \le \phi < 180^\circ$ the angular velocity $\Omega$ is negative and $v_{\rm rot}$ is blue-shifted.

\begin{figure*}[htb]
\plotone{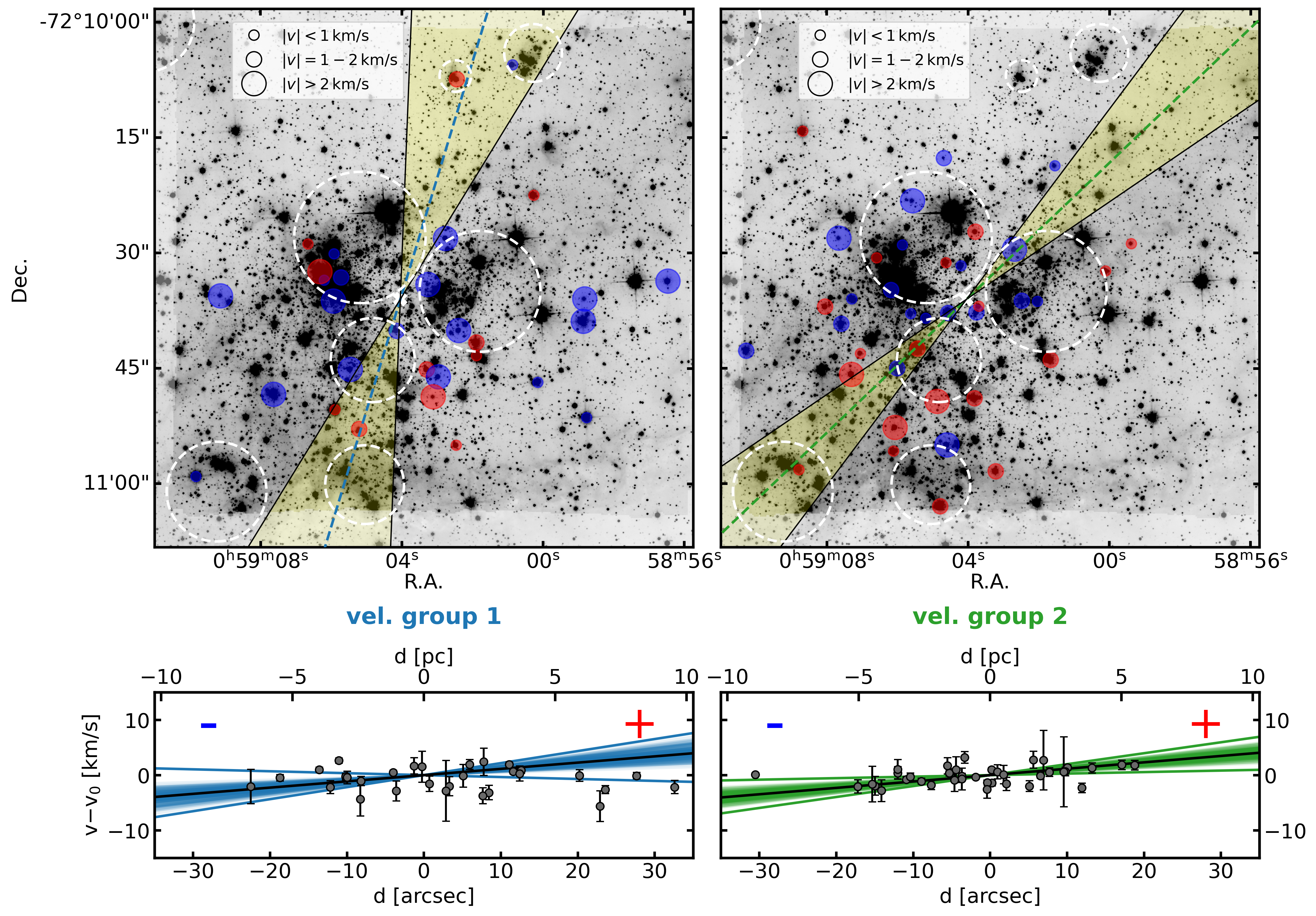}
\caption{This plot is similar to Fig.~\ref{fig:rot_all} but now for velocity groups 1 (left column) and 2 (right column). The top panel shows whether a star is red or blue shifted relative to the mean velocity of each group (see Tab.~\ref{tab:rv_stars}). The marker size indicates the magnitude of the velocity in three bins: $|v|<1\,{\rm km}/{\rm s}$, $1\,{\rm km}/{\rm s} \le |v| < 2\,{\rm km}/{\rm s}$, $|v| \ge 2\,{\rm km}/{\rm s}$). The bottom panel shows the stellar radial velocity relative to the distance to the rotation axis. The black line is the best-fit model while in blue and green we show a randomized sub-sample of all the MCMC solutions. The thicker line represents the most extreme solution for the fit.}
\label{fig:rot_groups}
\end{figure*}

Both scenarios converged with the following results: For scenario 1 (see Fig.~\ref{fig:rot_all}) we got $\Phi = 151^\circ \pm 23^\circ$ with $\Omega = (-0.4 \pm 0.2)\,{\rm Myr}^{-1}$. For scenario 2 (see Fig.~\ref{fig:rot_groups}) we got $\Phi_1 = 163^\circ \pm 16^\circ$ with $\Omega_1 = (-0.4 \pm 0.1)\,{\rm Myr}^{-1}$ and $\Phi_2 = 134^\circ \pm 10^\circ$ with $\Omega_2 = (-0.4 \pm 0.1)\,{\rm Myr}^{-1}$. 

For scenario 2, which is shown in Fig.~\ref{fig:rot_groups}, we only plot stars that could be uniquely assigned to either of the two velocity groups. The top row shows all stars that belong to each rotation group plotted over the {\em HST} $F814W$ image color-coded by their red or blue-shift. The marker size indicates the magnitude of the radial velocity in three bins: $|v|<1\,{\rm km}/{\rm s}$, $1\,{\rm km}/{\rm s} \le |v| < 2\,{\rm km}/{\rm s}$, $|v| \ge 2\,{\rm km}/{\rm s}$. The rotation axes are shown as dashed lines and the error cones of $\Phi$ in yellow. The bottom row shows the stellar radial velocities relative to the mean of each group plotted against the distance $d$ to the rotation axis. By definition of the chosen coordinate system, $d$ is negative for $0^\circ \le \phi < 180^\circ$ (indicated by the minus sign in all plots). The black line is the best-fit rotation model while in blue and green we show a randomized sub-sample of all the MCMC solutions with the outermost, thicker line being the most extreme case.

A visual inspection of the individual MCMC solutions of group 1 (see Fig.~\ref{fig:rot_groups}) reveals the presence of solutions with reversed rotation direction. Furthermore the error budget for group 1 is considerably larger. We thus conclude that only velocity group 2 shows a clear sign of rotation (see Sect.~\ref{sec:discussion} for a detailed discussion). To probe the significance of the detected rotation we run a simulation with 5000 realizations to test the likelihood of recovering our measured rotation profile from a spatially randomized, three velocity group population. Each realization represents the same sample size, radial velocity, and uncertainties as our dataset. Of all simulations, only 33.6\% show rotation. Of those simulations that show rotation only 4.8\% exceed the angular velocity measured in our data, which makes 1.5\% of all 5000 simulations. Given these numbers the likelihood of detecting the rotation by chance is interpreted as small.

Leaving the center of rotation as a free parameter, in the previous analysis yields no converging solution. This is most likely due to an incomplete sample and insufficient information about the cluster shape, the latter mainly driven by the limited FoV. We therefore arbitrarily chose as center of rotation the mean coordinates in ${\rm R.A.}=0^{\rm h}59^{\rm m}04^{\rm s}.039$ and ${\rm Dec.}=-72^\circ10^{\rm m}36^{\rm s}.04$, which roughly coincides with the center of sc\,1-3 (see Fig.~\ref{fig:NGC346_rgb}). An attempt to fit the RV distribution of Sec.~\ref{sec:RV_analysis} together with the rotation model to introduce varying, nested group membership probabilities yield the same results within uncertainties, hence we choose the least complex model. The \citet{Sabbi2022} proper motion study using multi-epoch {\em HST} photometry yielded with $v_{\rm rot,pm}^{{\rm max}} = -3.2\,{\rm km}/{\rm s}$ a maximum rotational velocity comparable to this work, especially when taking into account anisotropies, incompleteness, and projection effects. Their center of rotation is located outside our FoV, which we were unable to use due to an unstable, non-converging fit. Nevertheless, we used the photometric center as determined by that {\em HST} study, which is $5.28\,$arcsec apart. Using the same NGC~346 distance modulus as of 60.4\,kpc \citep{Smith1992,Hilditch2005,Glatt2008,Lemasle2017} this translates to a projected distance of 1.55\,pc. We repeated our rotation analysis with this new center of rotation and the results are identical within uncertainties. Yet, given our small number of stars, this center leads to an asymmetric distribution of stars and to larger uncertainties so we decided to use the results determined with our center for any further analysis.

An independent, parallel study of NGC~346 by \citet{Sabbi2022} using {\em HST} proper motions of upper main sequence stars showed an inwards spiraling motion originating in the Northeast. At a truncation radius of 10--13\,pc, which is on a similar scale as our limited MUSE FoV, these results are in good agreement with a solid-body rotator. This truncation radius is in agreement with the location of the majority of the stars in our group 2. Our projected rotation velocities at 5\,pc and 10\,pc are $v_{\rm rot,RV}^{5\,{\rm pc}} = (-2.98\pm0.34)\,{\rm km}/{\rm s}$ and $v_{\rm rot, RV}^{10\,{\rm pc}} = (-3.95\pm0.67)\,{\rm km}/{\rm s}$, respectively. This is in agreement with the maximum \citet{Sabbi2022} rotation velocity of $v_{\rm rot,pm}^{{\rm max}} = -3.2\,{\rm km}/{\rm s}$ taking into account anisotropies, incompleteness, and projection effects. Furthermore, the green proper motion component of \citet{Sabbi2022} stretches from the Northwest to the Southeast similar to our radial velocity group 1. 

As demonstrated in Sect.~\ref{sec:gas_vel_profile} the best-fitting gas velocity profile shows, with two Gaussians, a similar shape as the stars. Logically, we attempted to fit the same rotation model to gas group 2. Although the fit for gas group 2 nominally converged, a rotation is with $\Omega_{\rm gas} = (-0.033 \pm 0.02)\,{\rm Myr}^{-1}$ practically non-existent. Since gas group 1 only covers the Southern ridge (see Fig.~\ref{fig:gas_groups_map}) fitting a rotation curve is not feasible. Hence we conclude that the gas, at least on the scale of the MUSE FoV is not rotating. 

\section{Discussion and Conclusion}
\label{sec:discussion}

The decomposition of the stellar radial velocity profile of the central 1\,arcmin$^2$ of NGC~346 demonstrates that this cluster is built from multiple components that can only be distinguished by their kinematics. These results are confirmed by the independent proper motion study of \citet{Sabbi2022} using multi-epoch {\em HST} photometry. Multiple kinematic components have also been detected in other YSCs, yet it seems that their origin differs between individual star forming regions. In the ONC for example, \citet{Zari2019a} discovered kinematic sub-structures and suggested these are either the result of galactic shear or that it is the imprint of the parental GMC filaments, from which the sub-clusters had formed \citep{Fujii2021b}. The latter is similar to the conclusion drawn by \citet{Zeidler2021} for Wd2, another young Milky Way star cluster. This particular YSC shows five distinct radial velocity groups, of which two pairs belong to the two, coeval sub-clusters Wd2 is composed of and the fifth component is a halo-like structure. But compare to the ONC, it is believed that the onset of star formation in Wd~2 was triggered by the collision of at least two molecular clouds \citep{Furukawa2009,Furukawa2014,Ohama2010}.

\begin{figure}[htb]
\plotone{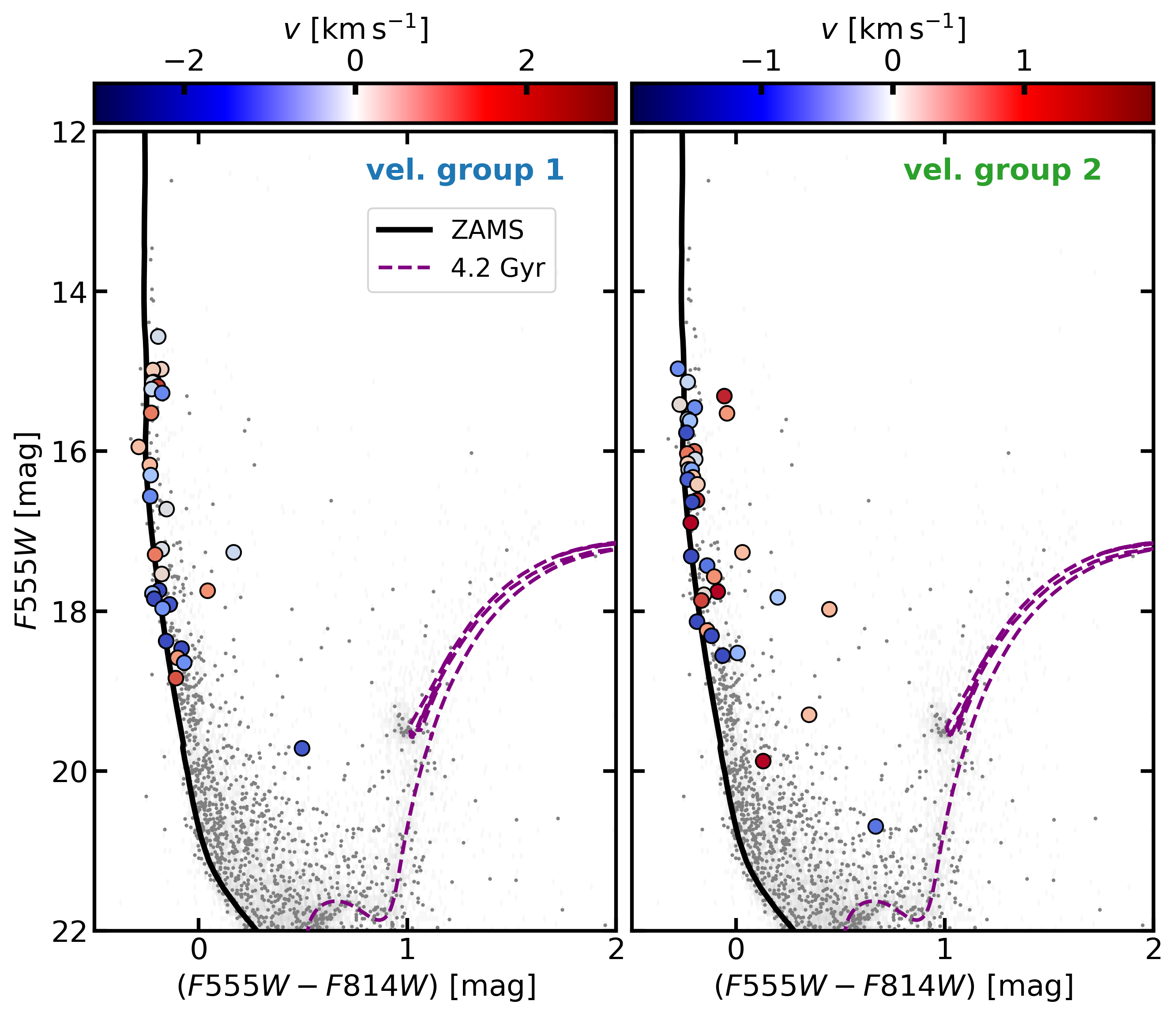}
\caption{The $F555W$ vs. $F555W-F814W$ CMD of the velocity groups 1 (left) and 2 (right). The stars are color-coded according to their velocity relative to each groups mean velocity. Similar to Fig.~\ref{fig:cmd} we show all {\em HST} detected stars in grey as well as the ZAMS and the 4.2\,Gyr isochrone representing the SMC field.}
\label{fig:cmd_vel_groups}
\end{figure}

Given that NGC~346 is also composed of multiple, coeval sub-clusters \citep{Sabbi2007} we started our analysis by trying to find a similar spatial correlation between individual sub-clusters and the two detected velocity groups. By comparing the stars' location for each velocity group (see Fig.~\ref{fig:rot_groups}), one can argue that velocity group 1 shows a more elongated spatial distribution in (North-)west -- (South-)east direction compared to velocity group 2, which is shaped rather spherical symmetric. Yet, this is not supported by a statistical analysis and in fact, with a $p$-value of 0.1 using a 2D Kolmogorov–Smirnov test \citep{Hodges1958,Peacock1983,Fasano1987} the null Hypothesis ``two individual velocity groups follow the same spatial distribution'' cannot be rejected. An inspection of the locus of these stars in the optical CMD (see Fig.~\ref{fig:cmd_vel_groups}) also shows no difference. Last but not least, we compared the stellar radial velocities of the individual sub-clusters \citep{Sabbi2007} and, within uncertainties, they are identical. Although, sub-clusters\,1--5, 7, and 8 are covered by the MUSE observations (see Fig.~\ref{fig:NGC346_rgb}), only sub-cluster\,1 -- 3 contain enough stars\footnote{sc\,1: 16, sc\,2: 10, sc\,3: 6, sc\,4: 1, sc\,5: 2, sc\,7: 1, sc\,8: 2,} with reliable velocity measurements. Hence, based on this dataset there is no direct correlation between the different sub-clusters and the kinematic components.

To further characterize the individual velocity groups we fit a rotation profile to the two kinematic components of NGC~346 using MCMC (see Sect.~\ref{sec:rot}). A thorough inspection of the posterior distributions showed that only velocity group 2 has a clear rotation signature. The larger scatter of the posterior distribution of velocity group 1 (see Fig.~\ref{fig:rot_groups}), where even solutions are possible for which the rotation direction changes, made the result rather ambiguous. That said, we are not excluding that rotation is also possible but the current MUSE dataset, especially the limited FoV, does not allow for a strong conclusion. The angular velocity of group 2 is with $\Omega_2 = (-0.4 \pm 0.1)\,{\rm Myr}^{-1}$ about half the value \citet{Henault-Brunet2012} found for R136 ($\Omega = (0.75 \pm 0.22)\,{\rm Myr}^{-1}$). Yet, given that the systems' inclinations are unknown and that R136 is more massive, which surely impacts its evolution (i.e., an increased gravitational potential) we conclude that these results are comparable. An independent, parallel study of NGC~346 by \citet{Sabbi2022} using {\em HST} proper motions of upper main sequence stars showed an inwards spiraling motion originating in the Northeast. At a truncation radius of 10--13\,pc, which is on a similar scale as our limited MUSE FoV, these results are in good agreement with a solid-body rotator. This truncation radius is in agreement with the location of the majority of the stars in our group 2. Our projected rotation velocities at 5\,pc and 10\,pc are $v_{\rm rot,RV}^{5\,{\rm pc}} = (-2.98\pm0.34)\,{\rm km}/{\rm s}$ and $v_{\rm rot, RV}^{10\,{\rm pc}} = (-3.95\pm0.67)\,{\rm km}/{\rm s}$, respectively. This is in agreement with the maximum \citet{Sabbi2022} rotation velocity of $v_{\rm rot,pm}^{{\rm max}} = -3.2\,{\rm km}/{\rm s}$ taking into account anisotropies, incompleteness, and projection effects. Furthermore, the green proper motion component of \citet{Sabbi2022} stretches from the Northwest to the Southeast similar to our radial velocity group 1. 

It is believed that such cluster rotation profiles can be produced by a hierarchical collapse of GMCs, where localized star formation events within this large structure leads to systematic accretion onto larger clumps along filament structures that represent the local gravitational potential \citep[e.g.,][and references therein]{Vazquez-Semadeni2019,Krause2020}. A recent study \citep{Dufton2019} of the most massive stars in N66 based on the spectral classification of VLT/FLAMES spectra found that stars in the innermost region, coinciding with the MUSE FoV, are more massive, may have lower projected rotational velocities\footnote{These are individual stellar rotations and not cluster rotations}, and are younger ($\le 2$\,Myr) compared to the overall age of $\sim3-6$\,Myr of NGC~346 \citep{Sabbi2008}. These findings support the argument of the global hierarchical collapse scenario, where the massive stars likely formed later than their lower-mass counterparts \citep[see Sect. 3.2 of][for a review on this topic]{Krause2020}. It is also important to mention that this scenario does not contradict \citet{Neelamkodan2021}, who proposed, based on Atacama Large Millimeter/sub-millimeter Array (ALMA) $^{12}{\rm CO}(1-0)$ data, that cloud-cloud collision is responsible for the on-going star formation in the region North-East of the cluster center (coinciding with Sc-13, see Fig.~\ref{fig:NGC346_rgb}). These cloud-cloud collisions are of small scale, compared to the N66 region and appear very localized. In fact, multiple class 0 young stellar objects found throughout NGC~346 at the interface of significant direction changes of the proper motion kinematic components \citep{Simon2007,Sewio2013,Rubio2018,Sabbi2022} and show that star formation is very well still on-going.

The analysis of the ionized gas in Sect.~\ref{sec:gas_vel_profile} revealed that, despite seeing two velocity peaks similar to the stellar radial velocity profile, no rotation can be detected. Furthermore, there is a strong spatial correlation of two gas velocity peaks (see Fig.~\ref{fig:gas_groups_map}), where the blue-shifted component coincides with the gas ridge south of the main star clusters. In velocity space, the stars that belong to velocity group 2 are located between the two gas peaks, essentially demonstrating an expanding gas bubble around the cluster stars; the red-shifted portion of the gas is located behind the stars and the blue-shifted component is pushed toward us. The visibility of the individual components depends on the LoS gas column density. In regions, such as the gas ridge toward the South, where the gas column density is high and the gas eventually becomes optically think, the blue-shifted component dominates. At locations where the sight line is rather perpendicular to the bubble's surface or where there is less material in front of the cluster (like toward the Northwest) we are able to ``see through'', hence measure the red-shifted component in the back of the cluster \citep[analogous to Westerlund~2,][]{Zeidler2021}. Due to the rather low pixel-to-pixel velocity resolution of MUSE and that velocity difference between the two peaks is only $7.9\,{\rm km}/{\rm s}$ a double-peaked line profile cannot be detected in regions where both components are visible, which was confirmed by a visual inspection. Given the large number of pixels and that we only draw qualitative conclusions the possible introduction of additional uncertainties due to broadened or skewed line profiles is minor. These results are in agreement with results based on Mopra and APEX CO data \citep{Muller2015}, as well as [\ion{C}{2}] data observed with SOFIA \citep{Requena-Torres2016}. Their velocity width of $\sim 20\,{\rm km}/{\rm s}$ resulting in an expansion velocity of 7--10\,${\rm km}/{\rm s}$ is in excellent agreement with our velocity range of the ionized gas (see Fig.~\ref{fig:rv_gas} and \ref{fig:rv_mod_gas}) and our measured gas peak to peak velocity of $7.9\,{\rm km}/{\rm s}$ (see Fig.~\ref{fig:rv_mod_gas}). \citet{Requena-Torres2016} also concluded that the bubble should have started expanding $\sim2$\,Myr ago, which is in excellent agreement with age of the massive stars in the cluster center \citep{Dufton2019}. The numerous massive stars of the central cluster photodissociated the molecular gas and their stellar winds and the far-ultraviolet fluxes started the expansion of this ionized gas bubble and have been driving it outwards ever since. Small differences in absolute velocities between the cold CO and the ionized gas is also seen in other regions such as the Lagoon Nebula \citep[M8,][]{Damiani2017} and might be caused by differences in the gas properties (e.g., feedback, sound speed).

\section{Summary}
\label{sec:summary}

In this work we present the radial velocity analysis of the central region of NGC~346 in the SMC using VLT/MUSE integral field spectroscopy. We utilize the high-precision astrometry and photometry from {\em HST} to extract stellar spectra to measure the radial velocities of 103 cluster member stars. With this established technique, implemented in \MUSEpack, we measure the velocities to an accuracy of $1.6\,{\rm km}/{\rm s}$, which allows us to study the cluster's internal stellar motion. Our main results are:

\begin{itemize}
    \item NGC~346 shows two distinct radial velocity groups at $v_1 = -3.3^{+0.3}_{-0.2}\,{\rm km}/{\rm s}$ and $v_2 = 2.6^{+0.1}_{-0.1}\,{\rm km}/{\rm s}$, and hints for a third group at $v_3 = 9.4^{+0.1}_{-0.1}\,{\rm km}/{\rm s}$, measured relative to the systemic velocity of $v_{\rm sys} = (165.5\pm0.2)\,{\rm km}/{\rm s}$. Taking into account projection effects these velocity groups are in agreement with the proper motion groups found by \citet{Sabbi2022} in an independent, parallel analysis.
    \item There is no significant correlation between the velocity groups, the stars' location or the sub-clusters. However, velocity group 1 appears to be slightly elongated in (North-)west -- (South-)east direction, while velocity group 2 is rather spherical symmetric (see Fig.~\ref{fig:rot_groups}). Also the stars of the two velocity groups are indistinguishable in optical CMDs (see Fig.~\ref{fig:cmd_vel_groups}).
    \item Velocity group 2 shows a clear rotation signal ($\Omega_2 = (-0.4 \pm 0.1)\,{\rm Myr}^{-1}$ or $v_{\rm rot,RV}^{10\,{\rm pc}} = (-3.9=\pm0.7)\,{\rm km}/{\rm s}$, see Fig.~\ref{fig:rot_groups}), which is comparable to the findings in R136 in the LMC \citep{Henault-Brunet2012}. \citet{Sabbi2022} found a similar rotation signature for the innermost region thus confirming our results. We conclude that this cluster rotation profiles is a result of the hierarchical collapse of the parental GMC that started star formation in N66.
    \item The ionizing gas of this central region shows clear signs of an expanding bubble. This expansion is caused and is driven by the stellar winds and ionizing fluxes of the many O and B stars and is in agreement with studies of the cold molecular gas and dust \citep{Muller2015,Requena-Torres2016}. This expansion erased any kinematic signature of the natal GMC.
\end{itemize}

This work demonstrates how powerful the combination of high-resolution {\em HST} photometry and MUSE integral field spectroscopy is to analyze the complicated kinematics of young star-forming regions, which allows us to separate and analyze individual stellar populations that are spatially co-located. It also shows that to fully understand the formation history of NGC~346 more observations are needed, especially from the spectroscopic side, to cover the full extent of the cluster region and to paint its the full 3D kinematic picture. Additionally, scheduled {\em James Webb Space Telescope} GTO observations will unveil and better characterize the youngest YSOs in this region, hence uncovering the precise locations of and to what extent star formation is still on-going.

\begin{acknowledgments}
We would like to express our gratitude to L.~Oskinova and W.-R. Hamann from the Institut f{\"u}r Physik und Astronomie of the Universit{\"a}t Potsdam for these excellent observations and for very productive and fruitful discussions, which we hope to continue in the future.

We thank M.~Libralato for all the clarifying discussion about rotation models including all the necessary coordinate transformations.

We thank the anonymous referee for their suggestions, which improved this work.

This work has made use of data from the European Space Agency (ESA) mission {\it Gaia} (\url{https://www.cosmos.esa.int/gaia}), processed by the {\it Gaia} Data Processing and Analysis Consortium (DPAC, \url{https://www.cosmos.esa.int/web/gaia/dpac/consortium}). Funding for the DPAC has been provided by national institutions, in particular the institutions participating in the {\it Gaia} Multilateral Agreement.
\end{acknowledgments}

\appendix

\section{Cluster member stars}
\label{sec:clm_stars}
In Tab.~\ref{tab:clm_stars} we list all 103 cluster member stars, for which we measured radial velocities.

\startlongtable
\begin{deluxetable}{rrrrrrrc}
	\tablecaption{The cluster member stars \label{tab:clm_stars}}
	\tablehead{\multicolumn{1}{c}{ID} & \multicolumn{1}{c}{R.A.} & \multicolumn{1}{c}{Dec.} & \multicolumn{1}{c}{$F555W$} & \multicolumn{1}{c}{$F814W$} & \multicolumn{1}{c}{RV} & \multicolumn{1}{c}{$\sigma$RV} & \multicolumn{1}{c}{Gaia ID}\\
	\multicolumn{1}{c}{} & \multicolumn{1}{c}{(ICRS)} & \multicolumn{1}{c}{(ICRS)} & \multicolumn{1}{c}{(mag)} & \multicolumn{1}{c}{(mag)} & \multicolumn{2}{c}{$({\rm km}/{\rm s}$)} & \multicolumn{1}{c}{}
		}
	\startdata
1 & 0$^{\rm h}$58$^{\rm m}$56.463$^{\rm s}$ & -72$^\circ$10$^{\rm m}$33.67$^{\rm s}$ & 16.559 & 16.793 & 160.1 & 1.2 & 4689015706360740480 \\
2 & 0$^{\rm h}$58$^{\rm m}$56.782$^{\rm s}$ & -72$^\circ$10$^{\rm m}$46.00$^{\rm s}$ & 22.445 & 22.086 & 19.5 & 4.1 & --- \\
3 & 0$^{\rm h}$58$^{\rm m}$57.272$^{\rm s}$ & -72$^\circ$10$^{\rm m}$28.81$^{\rm s}$ & 16.571 & 16.774 & 123.1 & 1.3 & 4689015706360736256 \\
4 & 0$^{\rm h}$58$^{\rm m}$57.368$^{\rm s}$ & -72$^\circ$10$^{\rm m}$33.67$^{\rm s}$ & 13.972 & 14.197 & 167.7 & 0.2 & 4689015706360740096 \\
5 & 0$^{\rm h}$58$^{\rm m}$58.766$^{\rm s}$ & -72$^\circ$10$^{\rm m}$51.39$^{\rm s}$ & 15.138 & 15.360 & 162.2 & 0.6 & 4689015706360755584 \\
6 & 0$^{\rm h}$58$^{\rm m}$58.823$^{\rm s}$ & -72$^\circ$10$^{\rm m}$35.97$^{\rm s}$ & 18.374 & 18.531 & 156.7 & 2.8 & 4689015702041058816 \\
7 & 0$^{\rm h}$58$^{\rm m}$58.862$^{\rm s}$ & -72$^\circ$10$^{\rm m}$38.87$^{\rm s}$ & 15.128 & 15.345 & 159.7 & 0.7 & 4689015706360742784 \\
8 & 0$^{\rm h}$58$^{\rm m}$59.201$^{\rm s}$ & -72$^\circ$11$^{\rm m}$01.42$^{\rm s}$ & 18.946 & 18.986 & 153.3 & 1.6 & 4689015702041061504 \\
9 & 0$^{\rm h}$58$^{\rm m}$59.376$^{\rm s}$ & -72$^\circ$10$^{\rm m}$28.79$^{\rm s}$ & 19.296 & 18.948 & 168.7 & 6.3 & 4689015706360734336 \\
10 & 0$^{\rm h}$58$^{\rm m}$59.471$^{\rm s}$ & -72$^\circ$10$^{\rm m}$45.26$^{\rm s}$ & 17.910 & 18.054 & 108.6 & 1.3 & 4689015702100221824 \\
11 & 0$^{\rm h}$59$^{\rm m}$00.111$^{\rm s}$ & -72$^\circ$10$^{\rm m}$32.33$^{\rm s}$ & 17.977 & 17.532 & 168.8 & 0.8 & 4689015702100196736 \\
12 & 0$^{\rm h}$59$^{\rm m}$00.151$^{\rm s}$ & -72$^\circ$10$^{\rm m}$46.83$^{\rm s}$ & 16.723 & 16.878 & 162.2 & 1.1 & 4689015706306943104 \\
13 & 0$^{\rm h}$59$^{\rm m}$00.188$^{\rm s}$ & -72$^\circ$10$^{\rm m}$03.32$^{\rm s}$ & 18.528 & 18.400 & 150.9 & 2.7 & 4689016462221084160 \\
14 & 0$^{\rm h}$59$^{\rm m}$00.265$^{\rm s}$ & -72$^\circ$10$^{\rm m}$22.50$^{\rm s}$ & 15.940 & 16.228 & 163.1 & 0.6 & 4689015706360729472 \\
15 & 0$^{\rm h}$59$^{\rm m}$00.743$^{\rm s}$ & -72$^\circ$10$^{\rm m}$28.16$^{\rm s}$ & 13.454 & 13.679 & 165.5 & 0.2 & 4689015702019200896 \\
16 & 0$^{\rm h}$59$^{\rm m}$00.866$^{\rm s}$ & -72$^\circ$10$^{\rm m}$05.54$^{\rm s}$ & 17.222 & 17.401 & 162.2 & 2.0 & 4689016458014441216 \\
17 & 0$^{\rm h}$59$^{\rm m}$01.000$^{\rm s}$ & -72$^\circ$10$^{\rm m}$16.25$^{\rm s}$ & 17.707 & 17.783 & 203.3 & 1.5 & 4689015702100223488 \\
18 & 0$^{\rm h}$59$^{\rm m}$01.109$^{\rm s}$ & -72$^\circ$10$^{\rm m}$22.56$^{\rm s}$ & 19.133 & 19.209 & 176.0 & 2.7 & 4689015706360699520 \\
19 & 0$^{\rm h}$59$^{\rm m}$01.544$^{\rm s}$ & -72$^\circ$10$^{\rm m}$18.68$^{\rm s}$ & 18.519 & 18.514 & 167.2 & 2.0 & 4689015706306867584 \\
20 & 0$^{\rm h}$59$^{\rm m}$01.665$^{\rm s}$ & -72$^\circ$10$^{\rm m}$43.92$^{\rm s}$ & 16.023 & 16.258 & 169.4 & 0.9 & 4689015706360722944 \\
21 & 0$^{\rm h}$59$^{\rm m}$01.799$^{\rm s}$ & -72$^\circ$10$^{\rm m}$31.23$^{\rm s}$ & 14.113 & 14.332 & 183.3 & 0.3 & 4689015706360700288 \\
22 & 0$^{\rm h}$59$^{\rm m}$01.885$^{\rm s}$ & -72$^\circ$10$^{\rm m}$43.35$^{\rm s}$ & 14.981 & 15.202 & 162.9 & 0.4 & 4689015706360723200 \\
23 & 0$^{\rm h}$59$^{\rm m}$01.888$^{\rm s}$ & -72$^\circ$10$^{\rm m}$41.65$^{\rm s}$ & 15.517 & 15.746 & 164.2 & 0.4 & 4689015706306929920 \\
24 & 0$^{\rm h}$59$^{\rm m}$01.904$^{\rm s}$ & -72$^\circ$10$^{\rm m}$21.49$^{\rm s}$ & 15.841 & 16.168 & 59.1 & 0.7 & 4689015702019202048 \\
25 & 0$^{\rm h}$59$^{\rm m}$01.938$^{\rm s}$ & -72$^\circ$10$^{\rm m}$11.78$^{\rm s}$ & 18.323 & 18.311 & 103.9 & 2.5 & 4689015702041075968 \\
26 & 0$^{\rm h}$59$^{\rm m}$02.039$^{\rm s}$ & -72$^\circ$10$^{\rm m}$36.31$^{\rm s}$ & 15.594 & 15.826 & 168.0 & 0.5 & 4689015706360710144 \\
27 & 0$^{\rm h}$59$^{\rm m}$02.388$^{\rm s}$ & -72$^\circ$10$^{\rm m}$40.09$^{\rm s}$ & 17.740 & 17.932 & 159.1 & 1.3 & 4689015706360703360 \\
28 & 0$^{\rm h}$59$^{\rm m}$02.442$^{\rm s}$ & -72$^\circ$10$^{\rm m}$07.43$^{\rm s}$ & 17.742 & 17.700 & 163.9 & 1.5 & 4689015740720425088 \\
29 & 0$^{\rm h}$59$^{\rm m}$02.461$^{\rm s}$ & -72$^\circ$10$^{\rm m}$55.02$^{\rm s}$ & 17.535 & 17.714 & 162.6 & 1.4 & 4689015637641280128 \\
30 & 0$^{\rm h}$59$^{\rm m}$02.474$^{\rm s}$ & -72$^\circ$10$^{\rm m}$36.23$^{\rm s}$ & 16.634 & 16.846 & 166.1 & 1.0 & 4689015706360717952 \\
31 & 0$^{\rm h}$59$^{\rm m}$02.684$^{\rm s}$ & -72$^\circ$10$^{\rm m}$29.51$^{\rm s}$ & 18.127 & 18.315 & 165.5 & 1.6 & 4689015706360710784 \\
32 & 0$^{\rm h}$59$^{\rm m}$02.761$^{\rm s}$ & -72$^\circ$10$^{\rm m}$28.12$^{\rm s}$ & 17.961 & 18.135 & 160.3 & 1.7 & 4689015702100225664 \\
33 & 0$^{\rm h}$59$^{\rm m}$02.843$^{\rm s}$ & -72$^\circ$10$^{\rm m}$37.47$^{\rm s}$ & 15.533 & 15.781 & 222.0 & 0.5 & 4689015706360706944 \\
34 & 0$^{\rm h}$59$^{\rm m}$02.908$^{\rm s}$ & -72$^\circ$10$^{\rm m}$34.90$^{\rm s}$ & 14.383 & 14.622 & 165.4 & 0.4 & 4689015706360713088 \\
35 & 0$^{\rm h}$59$^{\rm m}$02.966$^{\rm s}$ & -72$^\circ$10$^{\rm m}$46.11$^{\rm s}$ & 17.841 & 18.055 & 158.5 & 1.5 & 4689015706360713984 \\
36 & 0$^{\rm h}$59$^{\rm m}$02.974$^{\rm s}$ & -72$^\circ$10$^{\rm m}$33.08$^{\rm s}$ & 18.344 & 18.488 & 155.2 & 2.2 & 4689015706360716160 \\
37 & 0$^{\rm h}$59$^{\rm m}$03.048$^{\rm s}$ & -72$^\circ$10$^{\rm m}$44.17$^{\rm s}$ & 15.796 & 15.998 & 114.5 & 0.6 & 4689015706360713856 \\
38 & 0$^{\rm h}$59$^{\rm m}$03.112$^{\rm s}$ & -72$^\circ$10$^{\rm m}$48.72$^{\rm s}$ & 18.835 & 18.945 & 164.7 & 2.5 & 4689015706360749696 \\
39 & 0$^{\rm h}$59$^{\rm m}$03.215$^{\rm s}$ & -72$^\circ$10$^{\rm m}$58.39$^{\rm s}$ & 16.612 & 16.800 & 169.9 & 0.9 & 4689015633321605632 \\
40 & 0$^{\rm h}$59$^{\rm m}$03.258$^{\rm s}$ & -72$^\circ$10$^{\rm m}$34.12$^{\rm s}$ & 19.716 & 19.222 & 159.4 & 5.5 & 4689015706360716288 \\
41 & 0$^{\rm h}$59$^{\rm m}$03.293$^{\rm s}$ & -72$^\circ$10$^{\rm m}$45.16$^{\rm s}$ & 17.292 & 17.503 & 164.2 & 0.9 & 4689015706360705408 \\
42 & 0$^{\rm h}$59$^{\rm m}$03.651$^{\rm s}$ & -72$^\circ$10$^{\rm m}$48.82$^{\rm s}$ & 19.094 & 19.147 & 155.3 & 1.6 & 4689015706360701696 \\
43 & 0$^{\rm h}$59$^{\rm m}$03.696$^{\rm s}$ & -72$^\circ$10$^{\rm m}$36.94$^{\rm s}$ & 17.794 & 17.949 & 168.2 & 1.7 & 4689015706360715776 \\
44 & 0$^{\rm h}$59$^{\rm m}$03.760$^{\rm s}$ & -72$^\circ$10$^{\rm m}$37.73$^{\rm s}$ & 17.428 & 17.568 & 166.5 & 1.2 & 4689015706360715904 \\
45 & 0$^{\rm h}$59$^{\rm m}$03.791$^{\rm s}$ & -72$^\circ$10$^{\rm m}$27.25$^{\rm s}$ & 17.867 & 18.035 & 169.8 & 1.5 & 4689015740720445440 \\
46 & 0$^{\rm h}$59$^{\rm m}$03.813$^{\rm s}$ & -72$^\circ$10$^{\rm m}$48.91$^{\rm s}$ & 15.999 & 16.201 & 169.5 & 0.5 & 4689015637641224704 \\
47 & 0$^{\rm h}$59$^{\rm m}$03.893$^{\rm s}$ & -72$^\circ$10$^{\rm m}$22.23$^{\rm s}$ & 18.151 & 18.281 & 172.3 & 1.6 & 4689015740666619136 \\
48 & 0$^{\rm h}$59$^{\rm m}$03.961$^{\rm s}$ & -72$^\circ$10$^{\rm m}$51.16$^{\rm s}$ & 14.963 & 15.159 & 164.1 & 0.3 & 4689015637641238400 \\
49 & 0$^{\rm h}$59$^{\rm m}$04.122$^{\rm s}$ & -72$^\circ$10$^{\rm m}$50.17$^{\rm s}$ & 18.917 & 18.985 & 134.7 & 4.7 & 4689015637641238272 \\
50 & 0$^{\rm h}$59$^{\rm m}$04.146$^{\rm s}$ & -72$^\circ$10$^{\rm m}$40.14$^{\rm s}$ & 17.776 & 17.998 & 160.7 & 1.4 & 4689015740720463872 \\
51 & 0$^{\rm h}$59$^{\rm m}$04.201$^{\rm s}$ & -72$^\circ$10$^{\rm m}$31.66$^{\rm s}$ & 16.096 & 16.292 & 167.9 & 0.5 & 4689015740720453760 \\
52 & 0$^{\rm h}$59$^{\rm m}$04.225$^{\rm s}$ & -72$^\circ$10$^{\rm m}$25.48$^{\rm s}$ & 15.895 & 16.124 & 127.9 & 1.6 & 4689015736378943744 \\
53 & 0$^{\rm h}$59$^{\rm m}$04.260$^{\rm s}$ & -72$^\circ$10$^{\rm m}$27.24$^{\rm s}$ & 15.890 & 16.145 & 171.8 & 0.5 & 4689015740666623616 \\
54 & 0$^{\rm h}$59$^{\rm m}$04.359$^{\rm s}$ & -72$^\circ$10$^{\rm m}$14.90$^{\rm s}$ & 17.894 & 18.053 & 175.8 & 2.5 & 4689015736459966208 \\
55 & 0$^{\rm h}$59$^{\rm m}$04.432$^{\rm s}$ & -72$^\circ$10$^{\rm m}$45.45$^{\rm s}$ & 17.702 & 17.817 & 174.7 & 2.0 & 4689015740720440832 \\
56 & 0$^{\rm h}$59$^{\rm m}$04.479$^{\rm s}$ & -72$^\circ$10$^{\rm m}$24.77$^{\rm s}$ & 12.609 & 12.742 & 167.5 & 0.2 & 4689015740666629120 \\
57 & 0$^{\rm h}$59$^{\rm m}$04.535$^{\rm s}$ & -72$^\circ$10$^{\rm m}$48.52$^{\rm s}$ & 19.117 & 19.188 & 134.8 & 5.0 & 4689015633380814336 \\
58 & 0$^{\rm h}$59$^{\rm m}$04.567$^{\rm s}$ & -72$^\circ$10$^{\rm m}$37.82$^{\rm s}$ & 15.450 & 15.650 & 166.7 & 0.5 & 4689015740720455808 \\
59 & 0$^{\rm h}$59$^{\rm m}$04.601$^{\rm s}$ & -72$^\circ$10$^{\rm m}$54.97$^{\rm s}$ & 17.313 & 17.530 & 165.8 & 0.9 & 4689015637587484416 \\
60 & 0$^{\rm h}$59$^{\rm m}$04.626$^{\rm s}$ & -72$^\circ$10$^{\rm m}$31.27$^{\rm s}$ & 16.153 & 16.387 & 168.5 & 0.6 & 4689015740720460544 \\
61 & 0$^{\rm h}$59$^{\rm m}$04.683$^{\rm s}$ & -72$^\circ$10$^{\rm m}$17.66$^{\rm s}$ & 20.690 & 20.023 & 166.5 & 3.2 & 4689015740721130880 \\
62 & 0$^{\rm h}$59$^{\rm m}$04.788$^{\rm s}$ & -72$^\circ$11$^{\rm m}$02.96$^{\rm s}$ & 15.307 & 15.364 & 169.9 & 0.8 & 4689015633299729280 \\
63 & 0$^{\rm h}$59$^{\rm m}$04.877$^{\rm s}$ & -72$^\circ$10$^{\rm m}$49.30$^{\rm s}$ & 19.875 & 19.747 & 170.8 & 5.4 & 4689015667740665856 \\
64 & 0$^{\rm h}$59$^{\rm m}$05.194$^{\rm s}$ & -72$^\circ$10$^{\rm m}$38.51$^{\rm s}$ & 15.131 & 15.365 & 167.8 & 0.4 & 4689015740720467328 \\
65 & 0$^{\rm h}$59$^{\rm m}$05.208$^{\rm s}$ & -72$^\circ$10$^{\rm m}$52.90$^{\rm s}$ & 18.582 & 18.685 & 163.8 & 2.8 & 4689015672001014528 \\
66 & 0$^{\rm h}$59$^{\rm m}$05.435$^{\rm s}$ & -72$^\circ$10$^{\rm m}$42.43$^{\rm s}$ & 15.520 & 15.565 & 169.1 & 0.4 & 4689015740720451072 \\
67 & 0$^{\rm h}$59$^{\rm m}$05.459$^{\rm s}$ & -72$^\circ$10$^{\rm m}$45.17$^{\rm s}$ & 17.913 & 18.054 & 159.4 & 1.8 & 4689015672000989568 \\
68 & 0$^{\rm h}$59$^{\rm m}$05.574$^{\rm s}$ & -72$^\circ$10$^{\rm m}$23.23$^{\rm s}$ & 18.306 & 18.425 & 165.3 & 1.9 & 4689015736460033536 \\
69 & 0$^{\rm h}$59$^{\rm m}$05.623$^{\rm s}$ & -72$^\circ$10$^{\rm m}$37.91$^{\rm s}$ & 17.826 & 17.627 & 167.4 & 2.0 & 4689015740720463616 \\
70 & 0$^{\rm h}$59$^{\rm m}$05.715$^{\rm s}$ & -72$^\circ$10$^{\rm m}$33.17$^{\rm s}$ & 16.298 & 16.529 & 161.2 & 0.9 & 4689015740720458624 \\
71 & 0$^{\rm h}$59$^{\rm m}$05.861$^{\rm s}$ & -72$^\circ$10$^{\rm m}$28.95$^{\rm s}$ & 15.614 & 15.836 & 167.3 & 0.5 & 4689015740720438528 \\
72 & 0$^{\rm h}$59$^{\rm m}$05.894$^{\rm s}$ & -72$^\circ$10$^{\rm m}$50.37$^{\rm s}$ & 14.971 & 15.152 & 162.8 & 0.4 & 4689015671947195264 \\
73 & 0$^{\rm h}$59$^{\rm m}$05.922$^{\rm s}$ & -72$^\circ$10$^{\rm m}$30.11$^{\rm s}$ & 17.260 & 17.093 & 161.9 & 1.1 & 4689015740720443136 \\
74 & 0$^{\rm h}$59$^{\rm m}$05.939$^{\rm s}$ & -72$^\circ$10$^{\rm m}$36.27$^{\rm s}$ & 18.466 & 18.549 & 157.9 & 3.0 & 4689015740726068352 \\
75 & 0$^{\rm h}$59$^{\rm m}$06.004$^{\rm s}$ & -72$^\circ$10$^{\rm m}$44.99$^{\rm s}$ & 14.963 & 15.243 & 166.7 & 0.6 & 4689015672000962048 \\
76 & 0$^{\rm h}$59$^{\rm m}$06.069$^{\rm s}$ & -72$^\circ$10$^{\rm m}$52.71$^{\rm s}$ & 17.752 & 17.842 & 170.9 & 1.5 & 4689015672001014144 \\
77 & 0$^{\rm h}$59$^{\rm m}$06.110$^{\rm s}$ & -72$^\circ$10$^{\rm m}$55.78$^{\rm s}$ & 16.324 & 16.532 & 168.6 & 0.8 & 4689015671947215104 \\
78 & 0$^{\rm h}$59$^{\rm m}$06.176$^{\rm s}$ & -72$^\circ$10$^{\rm m}$34.82$^{\rm s}$ & 16.353 & 16.586 & 166.3 & 0.8 & 4689015740720446976 \\
79 & 0$^{\rm h}$59$^{\rm m}$06.193$^{\rm s}$ & -72$^\circ$10$^{\rm m}$33.56$^{\rm s}$ & 14.562 & 14.757 & 162.0 & 0.6 & 4689015740720446464 \\
80 & 0$^{\rm h}$59$^{\rm m}$06.246$^{\rm s}$ & -72$^\circ$10$^{\rm m}$36.64$^{\rm s}$ & 18.646 & 18.771 & 174.4 & 1.7 & 4689015740666614912 \\
81 & 0$^{\rm h}$59$^{\rm m}$06.316$^{\rm s}$ & -72$^\circ$10$^{\rm m}$32.41$^{\rm s}$ & 15.185 & 15.382 & 164.9 & 0.5 & 4689015736459927424 \\
82 & 0$^{\rm h}$59$^{\rm m}$06.589$^{\rm s}$ & -72$^\circ$10$^{\rm m}$30.62$^{\rm s}$ & 16.409 & 16.595 & 168.5 & 1.1 & --- \\
83 & 0$^{\rm h}$59$^{\rm m}$06.658$^{\rm s}$ & -72$^\circ$10$^{\rm m}$28.84$^{\rm s}$ & 16.168 & 16.403 & 163.2 & 0.4 & 4689015736459929472 \\
84 & 0$^{\rm h}$59$^{\rm m}$06.732$^{\rm s}$ & -72$^\circ$10$^{\rm m}$41.30$^{\rm s}$ & 14.470 & 14.672 & 169.8 & 0.3 & 4689015672000982912 \\
85 & 0$^{\rm h}$59$^{\rm m}$06.902$^{\rm s}$ & -72$^\circ$10$^{\rm m}$15.22$^{\rm s}$ & 18.368 & 18.474 & 112.3 & 2.2 & 4689015736459970816 \\
86 & 0$^{\rm h}$59$^{\rm m}$07.051$^{\rm s}$ & -72$^\circ$10$^{\rm m}$43.09$^{\rm s}$ & 18.240 & 18.379 & 169.1 & 2.3 & 4689015671947184000 \\
87 & 0$^{\rm h}$59$^{\rm m}$07.111$^{\rm s}$ & -72$^\circ$10$^{\rm m}$37.67$^{\rm s}$ & 18.674 & 18.772 & 136.6 & 3.0 & 4689015671947168640\\
88 & 0$^{\rm h}$59$^{\rm m}$07.292$^{\rm s}$ & -72$^\circ$10$^{\rm m}$35.96$^{\rm s}$ & 16.224 & 16.453 & 167.8 & 0.7 & 4689015672000971392 \\
89 & 0$^{\rm h}$59$^{\rm m}$07.303$^{\rm s}$ & -72$^\circ$10$^{\rm m}$45.79$^{\rm s}$ & 16.893 & 17.112 & 171.3 & 1.0 & 4689015667740461184 \\
90 & 0$^{\rm h}$59$^{\rm m}$07.352$^{\rm s}$ & -72$^\circ$10$^{\rm m}$42.65$^{\rm s}$ & 20.106 & 19.411 & 121.4 & 5.1 & 4689015672000985088 \\
91 & 0$^{\rm h}$59$^{\rm m}$07.416$^{\rm s}$ & -72$^\circ$10$^{\rm m}$12.91$^{\rm s}$ & 17.778 & 17.814 & 175.6 & 1.8 & 4689015736400844032 \\
92 & 0$^{\rm h}$59$^{\rm m}$07.589$^{\rm s}$ & -72$^\circ$10$^{\rm m}$39.22$^{\rm s}$ & 16.231 & 16.444 & 167.0 & 0.5 & 4689015672000988544 \\
93 & 0$^{\rm h}$59$^{\rm m}$07.627$^{\rm s}$ & -72$^\circ$10$^{\rm m}$48.37$^{\rm s}$ & 15.267 & 15.444 & 160.1 & 1.2 & 4689015667659471360 \\
94 & 0$^{\rm h}$59$^{\rm m}$07.654$^{\rm s}$ & -72$^\circ$10$^{\rm m}$28.10$^{\rm s}$ & 18.555 & 18.620 & 166.0 & 1.2 & 4689015740666608128 \\
95 & 0$^{\rm h}$59$^{\rm m}$08.045$^{\rm s}$ & -72$^\circ$10$^{\rm m}$36.99$^{\rm s}$ & 17.565 & 17.672 & 169.2 & 1.7 & 4689015672000981120 \\
96 & 0$^{\rm h}$59$^{\rm m}$08.143$^{\rm s}$ & -72$^\circ$10$^{\rm m}$32.85$^{\rm s}$ & 19.147 & 19.201 & 143.3 & 2.6 & 4689015667740676864 \\
97 & 0$^{\rm h}$59$^{\rm m}$08.504$^{\rm s}$ & -72$^\circ$10$^{\rm m}$22.07$^{\rm s}$ & 18.696 & 18.793 & 173.4 & 1.6 & 4689015736460041088 \\
98 & 0$^{\rm h}$59$^{\rm m}$08.586$^{\rm s}$ & -72$^\circ$10$^{\rm m}$06.83$^{\rm s}$ & 21.316 & 20.954 & 53.7 & 3.0 & --- \\
99 & 0$^{\rm h}$59$^{\rm m}$08.680$^{\rm s}$ & -72$^\circ$10$^{\rm m}$14.18$^{\rm s}$ & 15.410 & 15.681 & 168.2 & 0.3 & 4689015736378949504 \\
100 & 0$^{\rm h}$59$^{\rm m}$08.790$^{\rm s}$ & -72$^\circ$10$^{\rm m}$58.18$^{\rm s}$ & 17.262 & 17.232 & 168.7 & 1.3 & 4689015671947211904 \\
101 & 0$^{\rm h}$59$^{\rm m}$09.133$^{\rm s}$ & -72$^\circ$10$^{\rm m}$35.57$^{\rm s}$ & 18.644 & 18.714 & 160.2 & 3.1 & 4689015671947166976 \\
102 & 0$^{\rm h}$59$^{\rm m}$09.829$^{\rm s}$ & -72$^\circ$10$^{\rm m}$59.08$^{\rm s}$ & 15.216 & 15.444 & 161.8 & 0.6 & 4689015667659473920 \\
103 & 0$^{\rm h}$59$^{\rm m}$10.283$^{\rm s}$ & -72$^\circ$10$^{\rm m}$42.73$^{\rm s}$ & 15.765 & 16.005 & 166.2 & 1.7 & 4689015667659474176 \\
	\enddata
	\tablecomments{The 103 cluster member stars, for which we measured accurate radial velocities. We show an internal ID (Column 1), the coordinates of each star (Columns 2 and 3) and the {\em HST} $F555W$ and $F814W$ magnitudes (Columns 4 and 5) based on the \citet{Sabbi2007} photometry, the measured radial velocities and uncertainties (Columns 6 and 7), and the Gaia eDR3 \citepalias{GaiaCollaboration2020a}. Star 2 is not included in Gaia because it is too faint. While stars 82 and 98 are detected by Gaia, multiple {\em Hubble} sources are located at their coordinates, hence a unique match was not found. Star 87 is part of the Gaia catalog without valid photometry. We added its Gaia identifier manually for completeness.}
\end{deluxetable}

\bibliography{NGC346_MUSE}{}
\bibliographystyle{aasjournal}

\end{document}